\documentclass[onecolumn,showpacs,preprintnumbers,amsmath,amssymb]{revtex4}

\usepackage{graphicx}
\usepackage{rotating}
\usepackage{bm}
\usepackage{bbm}
\usepackage{ulem} 
\usepackage[usenames]{color}

\newcommand{\be}{\begin{eqnarray}}
\newcommand{\ee}{\end{eqnarray}}
\newcommand{\non}{\nonumber\\}

\newcommand{\lera}[1]{\left( #1 \right)}
\newcommand{\eqq}[1]{Eq. (\ref{#1})}
\newcommand{\ccc}[1]{Ref. \cite{#1}}

\begin{document}


\title{The $s$-wave pion-nucleus optical potential}

\author{M. \surname{D\"oring}}
\email{doering@ific.uv.es}
\affiliation{Departamento de F\'{\i}sica Te\'orica and IFIC,
Centro Mixto Universidad de Valencia-CSIC,\\
Institutos de
Investigaci\'on de Paterna, Aptd. 22085, 46071 Valencia, Spain}
\author{E. \surname{Oset}}
\email{oset@ific.uv.es}
\affiliation{Departamento de F\'{\i}sica Te\'orica and IFIC,
Centro Mixto Universidad de Valencia-CSIC,\\
Institutos de
Investigaci\'on de Paterna, Aptd. 22085, 46071 Valencia, Spain}


\begin{abstract}
We calculate the $s$-wave part of the pion-nucleus optical potential using a unitarized chiral approach that has been previously used to simultaneously describe pionic hydrogen and deuterium data as well as low energy $\pi N$ scattering  in the vacuum. This energy dependent model allows for additional isoscalar parts in the potential from multiple rescattering. We consider Pauli blocking and pion polarization in an asymmetric nuclear matter environment. Also, higher order corrections of the $\pi N$ amplitude are included. The model can accommodate the repulsion required by phenomenological fits, though the theoretical uncertainties are bigger than previously thought. At the same time, we also find an enhancement of the isovector part compatible with empirical determinations.
\end{abstract}
\pacs{%
36.10.Gv, 
24.10.Cn
}
\maketitle

\section{Introduction}
The problem of the missing repulsion in pionic atoms has attracted much attention in the past \cite{Ericson:1966fm,Thomas:1979xu,miller1,miller2, Garcia-Recio:1987ik,Vicente:1988iv,Garcia-Recio:1988fg,Migdal:1990vm,Oset:1995nr,Salcedo:1995ac} and recently \cite{Batty:1997zp,Kaiser:2001bx,Weise:2001sg, Kolomeitsev:2002gc,Chanfray:2003js,Girlanda:2003cq,Friedman:2007qx} and was further motivated by the discovery of deeply bound pionic atoms at GSI \cite{Yamazaki:1996zb, Yamazaki:1997na,Itahashi:1999qb,Gilg:1999qa}. 

Due to the repulsion of the $s$-state pion in nuclear matter, the $\pi^-$ wave function is strongly repelled and overlaps only little with the nucleus. The wave function tests mainly the peripheral zone of the nucleus and, thus, nuclear matter at less than nuclear density. However, even at half the nuclear matter density difficulties in the theoretical description persist. From phenomenological  fits to pionic atoms reaching from ${\rm C}$ to ${\rm Pb}$, a strong repulsion is needed for a consistent description of the combined data. However, theoretical calculations consistently failed to deliver this ''missing repulsion'' (see, e.g., \cite{Garcia-Recio:1987ik}) although there has been recent progress \cite{Kolomeitsev:2002gc}.

The $s$-wave pion-nucleus optical potential is the basic input for a calculation of the $s$-levels of pionic atoms. Usually, the optical potential is calculated for infinite nuclear matter as a function of the Fermi momentum. Explicit calculations for finite nuclei done in \ccc{Nieves:1993ev,Nieves:1991ye} provide a prescription to pass from nuclear matter to finite nuclei: the $s$-wave part of the potential is provided by the corresponding nuclear matter results changing $\rho$ to $\rho(r)$ (local density approximation), while for the $p$-wave the prescription is slightly more complicated. 

The $s$-wave pion optical potential $2\omega V_{\rm opt}(r)=\Pi_S(r)$ is closely connected to the $s$-wave pion selfenergy which is usually \cite{Ericson:1966fm}  parametrized as
\be
\Pi_S(r)=-4\pi\left[\left(1+\frac{m_\pi}{m_N}\right)b_0(\rho_p+\rho_n)
+\left(1+\frac{m_\pi}{m_N}\right)b_1 (\rho_n-\rho_p)
+\left(1+\frac{m_\pi}{2m_N}\right)B_0(\rho) (\rho_p+\rho_n)^2\right]
\label{param_opt}
\ee
where the density $\rho$ is a function of the radial distance, $\rho\equiv\rho(r)$, given by the density profile of the nucleus. From this expression the sensitivity of the selfenergy to the isoscalar $b_0$ becomes visible, as in symmetric nuclear matter the isovector term $b_1$ vanishes. However, heavy nuclei such as ${}^{208}_{82}{\rm Pb}$ recently used in experiments \cite{Yamazaki:1997na,Itahashi:1999qb,Gilg:1999qa} contain more neutrons than protons; it is therefore interesting to study asymmetric matter, in particular with respect to a possible renormalization of the isovector $b_1$ \cite{Oller:2001sn,Meissner:2001gz,Chanfray:2003js,Suzuki:2002ae}. The last term in Eq. (\ref{param_opt}) takes into account corrections from higher order in density. This quantity has also an imaginary part due to pion absorption, which is mainly a two-body process, and the imaginary part of the optical potential determines the width of the pionic atom. 

Traditional fits to pionic atoms \cite{Tauscher:1974bx,Stricker:1980vm,Nieves:1993ev} provide the set of parameters displayed in Tab. \ref{tab:bobo}.
\linespread{1.1}
\begin{table}[h]
\caption{Typical fits of pionic atom data.}
\begin{tabular*}{0.6\textwidth}{@{\extracolsep{\fill}}llll}
Ref. &$b_0\;[m_\pi^{-1}]$&$b_1\;[m_\pi^{-1}]$&$B_0\;[m_\pi^{-4}]$
\\ \hline \hline 
\cite{Tauscher:1974bx}&$-0.0045$&$-0.0873$&$-0.049+i\;0.046$
\\ 
\cite{Stricker:1980vm}&$-0.0325$&$-0.0947$&$0.002+i\;0.047$
\\
\cite{Nieves:1993ev}&$-0.0183$&$-0.105$&$i\;0.0434$
\\
\hline
\end{tabular*}
\label{tab:bobo}
\end{table}
\linespread{1.0}
Although the sets of parameters are quite different from each other they result in similar pion self energies at $\rho=\rho_0/2$, half the nuclear density.
Therefore these sets are not contradictory but tell us that the pionic atom data require this magnitude of selfenergy at $\rho_0/2$. This equivalence of pion optical potentials using the concept of $\rho_{\rm eff}=\rho_0/2$ was early established in \cite{Seki:1983sh,Seki:1983si}. Furthermore, Tab. \ref{tab:bobo} suggests that the smaller value of $|b_0|$ in Ref. \cite{Tauscher:1974bx} needs to be compensated by a large negative real part of the $\rho^2$-term $B_0$; thus, corrections of higher order in the density are important. 

The model of Ref. \cite{Doring:2004kt} is of interest in this context as a good part of the $\pi N$ vacuum isoscalar is generated by the multiple rescattering of the dominant Weinberg-Tomozawa term of the $\pi N$ interaction of isovector character. This realization is important because rescattering terms are appreciably modified in the nuclear medium. Indeed, the Pauli  blocking in the intermediate nucleon states is well known to generate a repulsion, the Ericson-Ericson Pauli corrected rescattering term \cite{Ericson:1966fm}. On the other hand, the pion polarization due to particle-hole ($ph$) and $\Delta$-hole ($\Delta h$) excitation of the intermediate pions also produces corrections and accounts for the imaginary part of the potential from pion absorption \cite{Oset:1979bi,Garcia-Recio:1987ik}. 

Another point is the energy dependence of the $\pi N$ interaction \cite{Kolomeitsev:2002gc}. Ref. \cite{Doring:2004kt} focuses on the precise determination of the scattering lengths but also provides the energy dependence close to threshold. For pionic atoms where the pion is practically at rest with respect to the nucleus this is still relevant due to the Fermi motion of the nucleons. Note that in this context, the vacuum model \cite{Doring:2004kt} already contains certain information about the nucleon-nucleon correlations as one of the fitted data points has been the $\pi^-$-deuteron scattering length. The deuteron wave function that enters the theoretical description provides the $NN$ momentum distribution and allows for an inclusion of the Fermi motion in the deuteron. The issue of the energy dependence is a relevant one and in the medium it induces corrections which, due to the smallness of the $b_0$ parameter, have an effect similar to a renormalization of $b_1$ \cite{Kolomeitsev:2002gc}.

On the other hand there are some medium corrections coming from vertex corrections, off-shell effects, and wave function renormalization which, if desired, can also be recast as renormalization of $b_1$ and $b_0$. We shall also introduce novel terms in the pion selfenergy related to the $N^*(1440)$ decay into $N\pi\pi$, with the two pions in a scalar isoscalar state. This mechanism has already been used in \cite{Meissner:2005bz} to estimate some uncertainties in the study of the $\pi$-deuteron interaction. 

Another novelty in the present work is that we shall start from a free model for $\pi N$ scattering which is constructed using a chiral unitary approach, incorporating the lowest order (LO), and the needed next to lowest order (NLO) chiral Lagrangians, together with multiple scattering of the pions \cite{Doring:2004kt}.

The vacuum model from Ref. \cite{Doring:2004kt} will be modified in various steps in the medium: In Sec. \ref{sec:full_model} Pauli blocking of the intermediate nucleonic states, together with the appropriate spectral function for the intermediate pions, will lead to non-linear corrections in the density with preliminary numerical results given in Sec. \ref{sec:numres}. Also, a self consistent calculation is presented in Sec. \ref{sec:selfcon} where the overall pion $s$-wave selfenergy serves as an input for the intermediate pions in the $\pi N$ loops. In Secs. \ref{sec:higher_order}, \ref{sec:renoiso}, the diagrammatic model will be extended to the above mentioned higher order vertex corrections. Final numerical results are provided in Sec. \ref{sec:numres2}.

\section{Low energy pion nucleon interaction in vacuum and matter}
\label{sec:vacuum}
The vacuum $\pi N$ isoscalar term $b_0$ is around ten times smaller than the vacuum isovector $b_1$-term  and its precise determination is a complex task due to large cancellations in the amplitude. With the advent of new experimental data \cite{Schroder:uq,Sigg:qd,Sigg:1995wb,Schroder:rc,Hauser:yd} for the  $\pi^-p\to\pi^-p$, $\pi^-p\to\pi^0p$, and $\pi^-d\to\pi^-d$ scattering lengths from pionic hydrogen and deuterium, theoretical efforts in several directions have been made to precisely determine the parameters of low energy $\pi N$ scattering. In this context, $\pi^-$-deuteron scattering at threshold plays an important role as the complex scattering length $a_{\pi^-d}$ puts tight constraints on the size of $b_0$. 

Pion deuteron scattering has been recently treated in chiral perturbative approaches \cite{Meissner:2005bz,Lensky:2006wd} including also corrections from isospin breaking \cite{Meissner:2005ne} and effects like Fermi motion \cite{Beane:2002wk}. These and other higher order corrections have been taken into account in another theoretical framework in \ccc{Ericson:2000md}. In the extraction of  the strong scattering lengths from experiment, special attention has to be paid to the Coulomb corrections in the extraction of the scattering lengths from pionic hydrogen \cite{Ericson:2004ts,Lyubovitskij:2000kk,Gasser:2002am}. 

\begin{figure}
\includegraphics[width=0.7\textwidth]{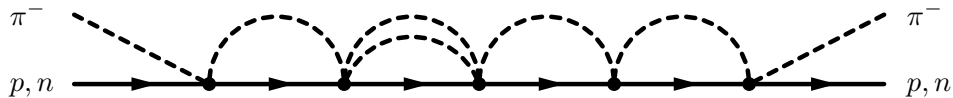}
\caption{Rescattering of the $\pi^- N$ system generated by the Bethe-Salpeter equation.}
\label{fig:bse}
\end{figure}

In the present study we rely upon a recent study on low energy $\pi N$ scattering in $s$-wave \cite{Doring:2004kt} which is summarized below. This model simultaneously describes the available data at threshold from pionic hydrogen and deuterium and also low energy $\pi N$ scattering. In a restriction to the coupled channels $\pi^-p,\, \pi^0 n$, and $\pi^-n$ the $\pi N$ $s$-wave amplitude $T(\sqrt{s})$ is unitarized by the use of the Bethe-Salpeter equation
\be
T(\sqrt{s})=\left[1-V(\sqrt{s})G(\sqrt{s}))\right]^{-1}V(\sqrt{s}).
\label{bsevacuum}
\ee
Here, the kernel $V$ is given by the elementary isovector interaction from the Weinberg-Tomozawa term of the LO chiral Lagrangian \cite{Meissner:1993ah,Bernard:1995dp,Ecker:1994gg},
\be
V_{ij}\left(\sqrt{s}\right)=-C_{ij}\;\frac{1}{4f_\pi^2}\;\left(2\sqrt{s}-M_i-M_j\right)\;\sqrt{\frac{M_i+E_i\left(\sqrt{s}\right)}{2M_i}}\;\sqrt{\frac{M_j+E_j\left(\sqrt{s}\right)}{2M_j}}.
\label{kernel_vac}
\ee
The $\pi N$ loop function $G$ in \eqq{bsevacuum} provides the unitarity cut and is regularized in dimensional regularization with one free parameter, the subtraction constant $\alpha_{\pi  N}$ \cite{Doring:2004kt}. In \eqq{kernel_vac} the coefficients $C_{ij}$ provide the transition strength of the coupled channels $i,j$ \cite{Doring:2004kt} and $M_{i,j}$, $E_{i,j}$ are the nucleon masses and energies. In Fig. \ref{fig:bse} we show a diagrammatic representation of the BSE equation (\ref{bsevacuum}), including also the $\pi\pi N$ channel which is also incorporated in \cite{Doring:2004kt}.

In the framework of the heavy baryon approach the vertices are factorized on-shell, see \eqq{kernel_vac}, because the off-shell part of the vertices in the loops can be absorbed renormalizing the lowest order tree level amplitude \cite{Oset:1997it}. However, we will see in Sec. \ref{sec:higher_order} that in a nuclear matter environment these renormalizations are modified leading to finite, density dependent corrections of the amplitude.

The multiple rescattering which is provided by Eq. (\ref{bsevacuum}) generates isoscalar pieces from the isovector interaction providing a large $b_0$ term. However, it is known \cite{Fettes:1998ud,Meissner:1999vr,Fettes:2000bb} that the NLO chiral Lagrangian is a necessary ingredient in $\pi N$ scattering at low energies. In order to provide the necessary degrees of freedom in the model, the isoscalar $s$-wave piece with the chiral coefficients $c_i$ in the notation of \ccc{Fettes:2000bb},
\be
V_{ij}\to V_{ij}+\delta_{ij} \left(\frac{4 c_1-2c_3}{f_\pi^2} \; m_\pi^2-2c_2\;\frac{(q^0)^2\;e^{-\beta^2\,(q^0)^2}}{f_\pi^2}\right)\frac{M_i+E_i\left(\sqrt{s}\right)}{2M_i},
\label{iso_lagrangian}
\ee
is added to the kernel of \eqq{bsevacuum}. The term $c_3\,q^2$ in ref. \cite{Fettes:2000bb} has been taken as $c_3\,m_\pi^2$, consistently with the approach of Refs. \cite{Inoue:2001ip,Oller:2000fj} which uses the on-shell values for the vertices in the scattering equations. The free fit parameters up to this point are the subtraction constant $\alpha_{\pi N}$ and the two combinations of $c_i$ from \eqq{iso_lagrangian}, as well as a damping factor parametrized with $\beta$ as discussed in Ref. \cite{Doring:2004kt} which is of no relevance here because we stay close to threshold. There are further refinements of the model, described in detail in Ref. \cite{Doring:2004kt}, such as the inclusion of the $\pi\pi N$ two-loop diagram which introduces one additional fit parameter, $\gamma$, from the small real part of this loop. 

In order to include the complex pion-deuteron scattering length $a_{\pi^-d}$ in the data fit, one has to employ the elementary $\pi N$ scattering model described above in the framework of a three body process. In \ccc{Doring:2004kt} this has been carried out by using the $\pi N$ amplitudes in a Faddeev multiple scattering approach. The interesting point is that the impulse approximation vanishes making the double rescattering off the two nucleons the dominant term. This term is sensitive to the isoscalar amplitude so that the experimental scattering length $a_{\pi^-d}$ provides valuable information on the vacuum $b_0$ term and sets tight constraints on it. 

Additional corrections of higher order \ccc{Ericson:2000md} in $\pi d$ scattering such as absorption, dispersion, the influence of the $\Delta(1232)$, and Fermi motion have been treated in a separate Feynman diagrammatic approach, together with other corrections from the literature, see \ccc{Ericson:2000md} and references therein. Once these various corrections are included in $a_{\pi d}$, the model parameters are fixed from data, namely the scattering lengths $a_{\pi^-p\to\pi^-p}$, $a_{\pi^-p\to\pi^0n}$, $a_{\pi^-d}$, and low energy $\pi N$ data from \cite{Arndt:2003if}. The parameter values are quoted in the left column of Tab. \ref{tab:parms}. The values of $c_i$ from Eq. (\ref{iso_lagrangian}) are in agreement with other works \cite{Fettes:2000bb}; furthermore, the isospin violations found in the study qualitatively agree with \ccc{Gibbs:1995dm}. In the following, we concentrate on the in-medium modifications of the approach.

\subsection{The model in nuclear matter}
\label{sec:full_model}
\begin{figure}
\includegraphics[width=0.5\textwidth]{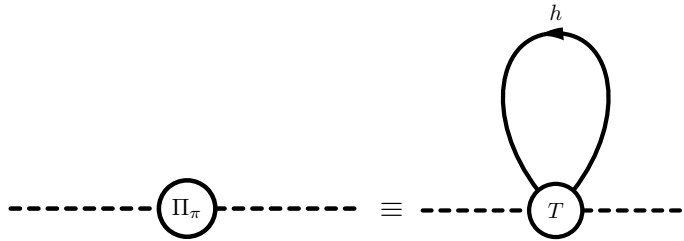}
\caption{Diagrammatic representation of the $\pi^-$ selfenergy from $s$-wave interaction with nucleons.}
\label{fig:symb_fermi}
\end{figure}

The $s$-wave $\pi N\to \pi N$ vacuum model from \ccc{Doring:2004kt}, summarized in Sec. \ref{sec:vacuum}, provides the driving interaction of the $\pi^-$ with the nucleus. In order to obtain the pion selfenergy $\Pi_S$ from \eqq{param_opt} of the $\pi^-$ in asymmetric nuclear matter with proton and neutron densities $\rho_p$ and $\rho_n$ ($k_F^p$, $k_F^n$ the respective Fermi momenta), the $\pi^- N\to \pi^- N$ amplitude $T$ is summed over the nucleons in the Fermi sea as schematically indicated in Fig. \ref{fig:symb_fermi}. The $s$-wave selfenergy for a $\pi^-$ at momentum $(k^0,{\bf k})$ with respect to the nuclear matter rest frame reads 
\be
\Pi_S(k^0, {\bf k};\rho_p,\rho_n)=2\int\limits^{k_F^{p}}\frac{d^3{{\bf p}_p}}{(2\pi)^3}\;T_{\pi^-p}(P^0,{\bf P};\rho_p, \rho_n)+2\int\limits^{k_F^{n}}\frac{d^3{{\bf p}_n}}{(2\pi)^3}\;T_{\pi^-n}(P^0,{\bf P};\rho_p,\rho_n)
\label{sum_fermi}
\ee
where ${\bf p}_{p,n}$ are the nucleon momenta. Due to the breaking of Lorentz invariance, the amplitudes  $T_{\pi^-p,n}$ depend independently on the components of $(P^0,{\bf P})$, the total $4$-momentum  of the $\pi N$ system in the nuclear matter frame, namely $P^0=k^0+E_{p,n}({\bf p}_{p,n})$ and ${\bf P}={\bf k}+{\bf p}_{p,n}$. The factors of 2 in Eq. (\ref{sum_fermi}) account for the sum over the nucleon spins. Note that Eq. (\ref{sum_fermi}) allows for isospin breaking by using different masses for particles of the same isospin multiplet.

In analogy to the vacuum case, $T_{\pi^-p}$ and $T_{\pi^-n}$ are given by the solutions of Bethe-Salpeter equations (BSE)
\be
T(P^0,{\bf P};\rho)=\left[1-V(\sqrt{s})G(P^0,{\bf P};\rho))\right]^{-1}V(\sqrt{s})
\label{bse}
\ee
where $s=(P^{0})^2-{\bf P}^2$ and the loop function $G$ is modified as described below. In Sec. \ref{sec:higher_order} we will apply in-medium changes also to the kernel $V$ from off-shell parts of the vertices and other sources. For the  charge $C=0$ sector, the BSE is represented by $(2\times 2)$ matrices accounting for the coupled channels $\pi^-p$ and $\pi^0n$. For the $\pi^- n$ interaction there is only one channel. 

The diagonal matrix $G$ from Eq. (\ref{bse}) contains the loop functions $G_{\pi N}$ which have been formulated in dimensional regularization in Ref. \cite{Doring:2004kt,Inoue:2001ip} for the vacuum case. Alternatively, one can use a cut-off scheme \cite{Inoue:2001ip} with $\Lambda$ the three momentum cut-off. The vacuum $G_{\pi N}$ is then given by
\be
G_{\pi N}(P^0, {\bf P})=a_{\pi N}+i\int\frac{d^4 q}{(2\pi)^4)}\frac{M_N}{E({\bf P}-{\bf q})}\;\frac{1}{P^0-q^0-E({\bf P}-{\bf q})+i\epsilon}\;\frac{1}{(q^0)^2-{\bf q}^2-m+i\epsilon}
\label{gcut}
\ee
with a cut-off for the three-momentum integration $\Lambda=$ 1 GeV and $m$ $(M_N)$ being the $\pi^-, \pi^0$ $(p,n)$ masses. Over wide energy ranges, a change in $\Lambda$ can be written as an additive constant to the real part of $G_{\pi N}$. Therefore, we have denoted a separate piece $a_{\pi N}$ in \eqq{gcut} in the same way as in Ref. \cite{Inoue:2001ip}. For the free case the propagator in the cut-off scheme agrees with the propagator from dimensional regularization over a wide energy range by choosing the appropriate subtraction constant. In the nuclear medium with Lorentz covariance explicitly broken, a cut-off scheme is more convenient in order to implement the in-medium dressing. Thus, we will employ the propagator from Eq. (\ref{gcut}) in this work. This requires a refit of the vacuum data. The values of the model parameters with the cut-off propagator from Eq. (\ref{gcut}) instead of dimensional regularization are displayed in Tab. \ref{tab:parms} on the right hand side. The new fit shows that the model is insensitive to the used regularization scheme. Parameters, $\chi^2$, and predictions for isoscalar and isovector terms $b_0$ and $b_1$ are stable. For notation of the parameters, see Sec. \ref{sec:vacuum}. In Tab. \ref{tab:parms}, $\alpha_{\pi N}$ is the subtraction constant of the loop in dimensional regularization and $a_{\pi N}$ the subtraction constant from \eqq{gcut}.
\linespread{1.1}
\begin{table}
\caption{Global fits to pionic hydrogen, deuteron, and low energy $\pi N$ scattering data, using dimensional regularization from Ref. \cite{Doring:2004kt} and cut-off scheme. Also, the resulting $b_0,b_1$ are shown.}
\begin{tabular*}{0.8\textwidth}{@{\extracolsep{\fill}}lll}
&{\bf DimReg}&{\bf Cut-off}
\\ \hline \hline
fitted data ($\sqrt{s}$)&1104--1253 MeV + threshold&1104--1253 MeV + threshold
\\ 
$\chi_r^2$&$51/(2\cdot 10+3)\simeq 2.2$&$48/(2\cdot 10+3)\simeq 2.1$
\\
$\alpha_{\pi N}$ [-]&$-1.143\pm 0.109$&---
\\
$a_{\pi N}$ [MeV]&---&$-2.025 \pm 1.28$
\\
$2c_1-c_3$ [GeV$^{-1}$]&$-1.539\pm 0.20$&$-1.487 \pm 0.20$
\\
$c_2$ [GeV$^{-1}$]&$-2.657\pm 0.22$&$-2.656 \pm 0.22$
\\
$\beta$ [MeV$^{-2}$]&$0.002741\pm 1.5\cdot 10^{-4}$&$0.002752\pm 1.5\cdot 10^{-4}$
\\
$\gamma\; [10^{-5}\cdot m_\pi^{5}]$&$5.53\pm 7.7$&$6.27 - 7.8$
\\ \hline 
$\chi^2(a_{\pi^- p\to \pi^-p})$&$3$&$3$
\\
$\chi^2(a_{\pi^- p\to \pi^0 n})$&$<1$&$<1$
\\
$\chi^2(a_{\pi^- d})$&$8$&$7$
\\ \hline
$b_0$ [$10^{-4}\;m_{\pi^-}^{-1}$]&$-28\pm 40$&$-29$
\\
$b_1$ [$10^{-4}\;m_{\pi^-}^{-1}$]&$-881\pm 48$&$-883$
\\ \hline
\end{tabular*}
\label{tab:parms}
\end{table}
\linespread{1.0}
The more important parameters are the $c_i$ and $\alpha_{\pi N} (a_{\pi N})$. The real part of the $\pi\pi N$ loop ($\gamma$) is tiny at threshold. For pionic atoms, the damping factor $\beta$ of \cite{Doring:2004kt} which is more important for the higher energy $\pi N$ data is of no relevance because the c.m. energy of $\pi N$ due to Fermi motion in the nucleus is small.

\begin{figure}
\includegraphics[width=0.4\textwidth]{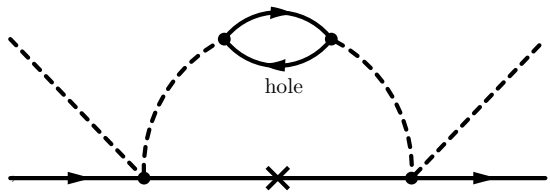}
\caption{In-medium correction of $s$-wave $\pi N$ scattering: Renormalization of the pion and Pauli blocking of the nucleon, symbolized by a crossed propagator. The pion $p$-wave selfenergy stands for resummed $ph$, $\Delta h$ insertions and includes $NN,\, N\Delta,\,\Delta\Delta$ short-range correlations.}
\label{fig:inmedium}
\end{figure}

The two major medium modifications of $G_{\pi N}$ are the Pauli blocking of the nucleon propagator and the polarization of the pion. The corresponding diagram is displayed in Fig. \ref{fig:inmedium}.
For the amplitude of the in-medium $\pi N$ loop function a similar expression as in Ref. \cite{Inoue:2002xw} is obtained. Here, we give the generalization to asymmetric nuclear matter for the $\pi^-p$, $\pi^0 n$, and $\pi^- n$ loops. With $N=p,n$ and $\pi_i=\pi^-, \pi^0$,
\be
&&G_{\pi_i N}(P^0, {\bf P};\rho_p,\rho_n)=a_{\pi N}+i\int\frac{d^4q}{(2\pi)^4}\theta(q_{\rm cm}^{\rm max}-|{\bf q}_{\rm cm}|)\frac{M_N}{E_N({\bf P-q})}\non
&\times&\left(\frac{\theta(|{\bf P-q}|-k_F^N)}{P^0-q^0-E_N({\bf P-q})+i\epsilon}+\frac{\theta(k_F^N-|{\bf P-q}|)}{P^0-q^0-E_N({\bf P-q})-i\epsilon}\right)\int\limits_0^\infty d\omega\;\frac{2\omega}{(q^0)^2-\omega^2+i\epsilon}\;S_{\pi_i}(\omega, {\bf q};\rho_p, \rho_n).
\label{matterg}
\ee
The cut-off in the vacuum model is applied in the $\pi N$ c.m. frame, as required by the vacuum model, whereas Eq. (\ref{matterg}) is defined in the nuclear matter rest frame. Since in the free case $q_{\rm max}$ is given in the c.m. frame we boost ${\bf q}$ to this frame and demand it to be smaller in modulus than $q_{\rm cm}^{\rm max}$. We have
\be
{\bf q}_{\rm cm}=\left[\left(\frac{P^0}{\sqrt{s}}-1\right)\frac{{\bf P\cdot q}}{|{\bf P}|^2}-\frac{q^0}{\sqrt{s}}\right]{\bf P}+{\bf q}
\label{cut-off}
\ee
where $s=(P^0)^2-{\bf P}^2$.
In Eq. (\ref{matterg}) we have also taken into account the hole part of the nucleon propagator as in Ref. \cite{Ramos:1999ku} which can play a role at the low pion energies we are studying. This term has been neglected in Ref. \cite{Inoue:2002xw} which is justified at higher energies. The pion spectral function $S_{\pi_i}$ is different for $\pi^-$ and $\pi^0$ for asymmetric nuclear matter. For $S$ we include the particle hole $(ph)$ excitation and $NN$ short-range correlations as described in the next section. 

In the model from Ref. \cite{Doring:2004kt}, the $\Delta(1232)$ has been explicitly taken into account in pion-deuteron scattering, leading to corrections in the $\pi d$ scattering length which by itself sets constraints on the vacuum isoscalar amplitude. In the present situation we can take the corresponding effect into account by including also the $\Delta$-hole ($\Delta h$) excitation in the pion selfenergy; in fact, closing the nucleon lines of the deuteron in the $\Delta$-box and $\Delta$-crossed box diagrams of Ref. \cite{Doring:2004kt} one obtains a pion selfenergy corresponding to Fig. \ref{fig:inmedium} substituting the $ph$ by a ${\Delta}h$ excitation of the pion. The $N^*(1440)$ Roper-hole excitation can be in principle also included in the pion selfenergy but has been found small in Ref. \cite{Inoue:2002xw} for low energy pions. However, in Sec. \ref{sec:roper} the Roper resonance will be included in a different context based on the coupling of the Roper to a scalar-isoscalar pion pair.

In Ref. \cite{Inoue:2002xw} Pauli blocking for the intermediate $\pi\pi N$ loop -- see Fig. \ref{fig:bse} -- has been included for the imaginary part. In the present case the pion has very little momentum in the $\pi N$ c.m. frame and the system is below the $\pi\pi N$ threshold, hence, even in vacuum the imaginary part of this term is zero and thus no change is required. Hence, the imaginary part of the amplitude from the $\pi\pi N$ intermediate state is zero in our case and the contribution to the real part of the amplitude is in any case negligible.

\begin{figure}
\includegraphics[width=0.4\textwidth]{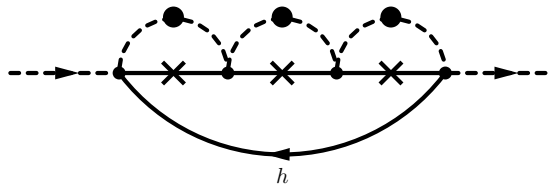}
\caption{Integration over the Fermi sea of the medium $\pi N$ amplitude. The crosses represent Pauli blocking of the nucleon propagators and the large dots, the $p$-wave pion selfenergy.}
\label{fig:inmedium_closed}
\end{figure}

Combining all the ingredients of the in-medium model, the $s$-wave pion selfenergy can be symbolized by the diagram in Fig. \ref{fig:inmedium_closed}: The in-medium propagator from Eq. (\ref{matterg}) is used in the BSE (\ref{bse}), and the remaining integral over the Fermi seas from Eq. (\ref{sum_fermi}) corresponds to closing the nucleon line. 

\subsection{Pion polarization in asymmetric nuclear matter}
\label{sec:pwaveself}
The spectral function of the pion $\pi_i$ ($\pi^+,\,\pi^-,\,\pi^0$) at momentum $(q^0,{\bf q})$ from Eq. (\ref{matterg}) is given by the imaginary part of the propagator, 
\be
S_{\pi_i}(q^0, {\bf q};\rho_p, \rho_n)&=&-\frac{1}{\pi}\;{\rm Im}\;D_{\pi_i},\quad D_{\pi_i}=\frac{1}{(q^0)^2-{\bf q}^2-m_{\pi_i}^2-
\Pi_{\pi_i}(q^0, {\bf q};\rho_p, \rho_n)}.
\label{spectral}
\ee
For the pion selfenergy inside loops the $p$-wave part is dominant because ${\bf q}$ is a running variable and $\Pi_{\pi_i}\varpropto {\bf q}^2$. The $s$-wave part will be included in the self consistent treatment in Sec. \ref{sec:selfcon}.
For the selfenergy we take into consideration the $(ph)-(ph)$ short-range repulsion parametrized in terms of the Migdal parameter, which is chosen $g'=0.7$, 
\be
\Pi_{\pi_i}(q^0, {\bf q};\rho_p, \rho_n)&=&\left(\frac{D+F}{2f_\pi}\right)^2\,F^2(q)\, {\bf q}^2\;\frac{U_{\pi_i}(q^0, {\bf q};\rho_p, \rho_n)}{1-\left(\frac{D+F}{2f_\pi}\right)^2\,F^2(q)\;g'\;U_{\pi_i}(q^0, {\bf q};\rho_p, \rho_n)}.
\label{migdal}
\ee
For a diagrammatic representation of the pion selfenergy, see, e.g., Ref. \cite{Oset:1995nr}. The Lindhard functions for asymmetric matter for $(ph)$ and $(\Delta h)$ excitations, evaluated below, are added in Eq. (\ref{migdal}), $U=U^{(ph)}+U^{(\Delta h)}$. Note that we have here for simplicity assigned the same $g'$ to $(ph)$ and $(\Delta h)$ excitations. For the form factor that takes into account the off-shell pions coupling to $ph$ or $\Delta h$ we have chosen the same function $F(q)=\Lambda^2/(\Lambda^2+{\bf q}^2)$ with $\Lambda=0.9$ GeV. 

The Lindhard function for symmetric nuclear matter, $U(q,k_F)$, can be found in the literature, e.g. in Ref. \cite{Oset:1989ey}, and here, we concentrate on an extension to asymmetric matter (see also \cite{Biswas:2007zs}). In the non-relativistic reduction, the Lindhard function for pions turns out to be 
\be
U_{\pi_i} (q,k_F^1, k_F^2)=4\int\frac{d^3{\bf k}}{(2\pi)^3}\;\left[\frac{\Theta(k_F^1-|{\bf k}|)\;\Theta(|{\bf k+q}|-k_F^2)}{q^0+\epsilon({\bf k})-\epsilon({\bf k+q})+i\eta}+\frac{\Theta(k_F^2-|{\bf k}|)\;\Theta(|{\bf k-q}|-k_F^1)}{-q^0+\epsilon({\bf k})-\epsilon({\bf k-q})+i\eta}\right].
\label{asym_u}
\ee
The first term is the contribution of the forward going $ph$ excitation (direct term) and the second term the pion crossed-term selfenergy. The index 1 (2) labels the Fermi sea corresponding to the hole (particle) part of the direct and the particle (hole) part of the crossed contribution. E.g., for a $\pi^-$, $k_F^1=k_F^p$ and $k_F^2=k_F^n$. For a $\pi^+$, $k_F^1=k_F^n$, $k_F^2=k_F^p$. The integral (\ref{asym_u}) can be solved analytically. For this, we split the ordinary Lindhard function from Ref. \cite{Oset:1989ey} in direct and crossed part by $U(q^0,{\bf q},k_F)=U_d(q^0,{\bf q},k_F)+U_c(q^0,{\bf q},k_F)$ with $U_c(q^0,{\bf q},k_F)=U_d(-q^0,{\bf q},k_F)$ and
\be
U_d(q^0,{\bf q},k_F)&=&\frac{3}{2}\;\frac{\rho M_N}{|{\bf q}|k_F}\left(z+\frac{1}{2}\left(1-z^2\right)\log\left(\frac{z+1}{z-1}\right)\right), \quad z=\frac{M_N}{|{\bf q}|k_F}\left(q^0-\frac{{\bf q}^2}{2M_N}\right)
\ee
where $\rho=2/(3\pi^2)k_F^3$ and $M_N$ the proton or neutron mass. Evaluating the integral in Eq. (\ref{asym_u}) one obtains for the $ph$ Lindhard function in asymmetric matter
\be
U^{(ph)}_{\pi^+}(q^0, {\bf q};\rho_p, \rho_n)&=&U_d(q^0,{\bf q},k_F^n)+U_c(q^0,{\bf q},k_F^p),\non
U^{(ph)}_{\pi^-}(q^0, {\bf q};\rho_p, \rho_n)&=&U_d(q^0,{\bf q},k_F^p)+U_c(q^0,{\bf q},k_F^n),\non
U^{(ph)}_{\pi^0}(q^0, {\bf q};\rho_p, \rho_n)&=&\frac{1}{2}\left(U(q^0,{\bf q},k_F^p)+U(q^0,{\bf q},k_F^n)\right).
\label{asym_u_sol}
\ee
These are the expressions to be used in Eq. (\ref{migdal}). The result in Eqs. (\ref{asym_u}) and (\ref{asym_u_sol}) is in agreement with Ref. \cite{Urban:2002ie}, correcting a typographical error in their Eq. (A.5).

For the $\Delta h$ Lindhard function $U^{(\Delta h)}(q,k_F^p, k_F^n)$, no new calculation is required, as the $\Delta$ always plays the role of a particle and is not affected by the Fermi sea. It is therefore sufficient to split $U_\Delta (k_F)$ from Ref. \cite{Oset:1989ey} into its charge states and direct plus crossed parts, and use as argument the $k_F$ that corresponds to the hole part, 
\be
U_{\pi^-}^{(\Delta h)}(q^0, {\bf q};\rho_p, \rho_n)&=&\frac{1}{4}U^{(\Delta h)}_d(q^0, {\bf q};k_F^p)+ \frac{3}{4}U^{(\Delta h)}_{c}(q^0, {\bf q};k_F^p)+\frac{1}{4}U^{(\Delta h)}_{c}(q^0, {\bf q};k_F^n)+  \frac{3}{4}U^{(\Delta h)}_{d}(q^0, {\bf q};k_F^n),\non
U_{\pi^0}^{(\Delta h)}(q^0, {\bf q};\rho_p, \rho_n)&=&\frac{1}{2}\left(U^{(\Delta h)}_d(q^0, {\bf q};k_F^p)+ U^{(\Delta h)}_{c}(q^0, {\bf q};k_F^p)+U^{(\Delta h)}_{c}(q^0, {\bf q};k_F^n)+U^{(\Delta h)}_{d}(q^0, {\bf q};k_F^n)\right).
\ee
Analytic expressions for the direct and crossed part of the $\Delta h$ Lindhard function can be found in Ref. \cite{Oset:1989ey}. 

\begin{figure}
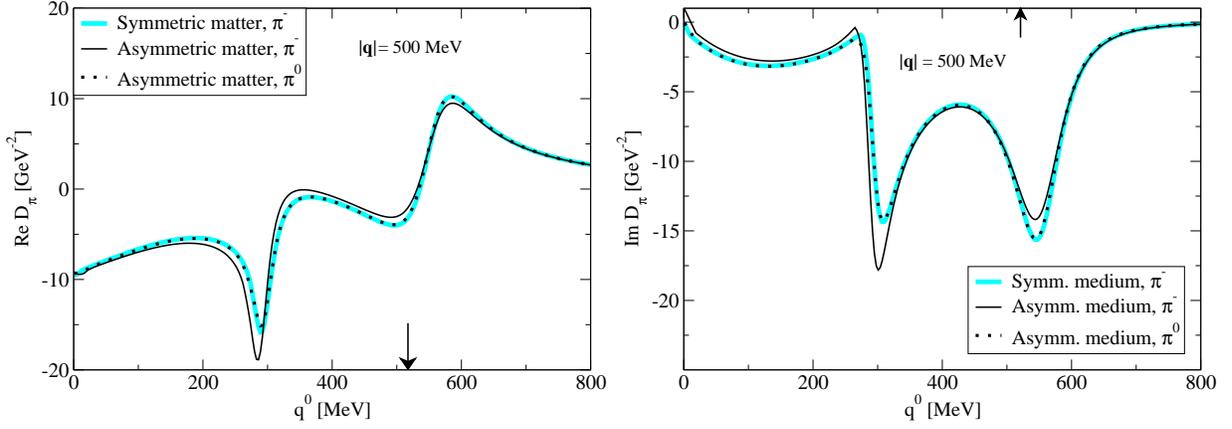

\includegraphics[width=8cm]{fig051_op.eps}
\includegraphics[width=8cm]{fig052_op.eps}
\caption{Real and imaginary part of the pion propagator $D_\pi$ for symmetric and asymmetric nuclear matter at a pion momentum of 500 MeV. The position of the quasielastic pion peak in vacuum, at $q^0=518$ MeV, is indicated with the arrows. The asymmetric matter corresponds to the ratio of $n$ to $p$ in ${}^{208}_{82}{\rm Pb}$.}
\label{fig:propagator}
\end{figure}

In order to see the effects of asymmetric nuclear matter we plot the pion propagator for normal nuclear density $\rho_0=0.483\;m_\pi^3$ which corresponds to $k_F=268$ MeV for symmetric matter. For asymmetric matter we set $k_F^n=1.154\;k_F^p$ which corresponds to the ratio of neutron rich nuclei such as ${}^{208}_{82}{\rm Pb}$. Then, $\rho_0=\rho_p+\rho_n$ is obtained with $k_F^p=247$ MeV and $k_F^n=286$ MeV. In the plots in Fig. \ref{fig:propagator} the propagator from Eq. (\ref{spectral}) for pions at $|{\bf q}|=500$ MeV is shown. 
The $\pi^0$ in asymmetric matter is very similar to the case of symmetric matter. The $\pi^-$ shows some minor deviations. 

\section{Numerical results}
\label{sec:numres}
\begin{figure}
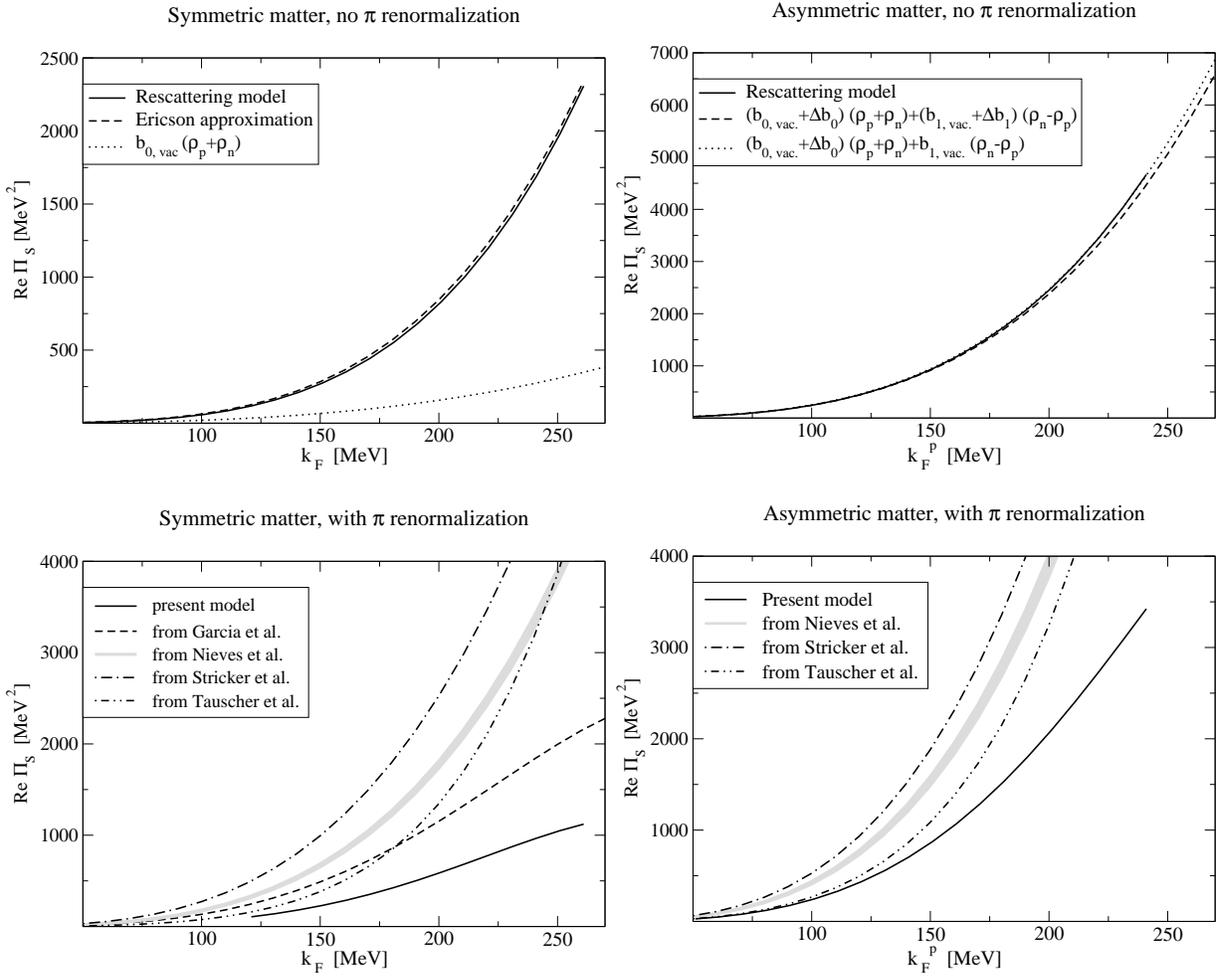

\includegraphics[width=8cm]{fig061_op.eps}
\includegraphics[width=8cm]{fig062_op.eps}\\ 
\vspace*{0.4cm}
\includegraphics[width=8cm]{fig063_op.eps}
\includegraphics[width=8cm]{fig064_op.eps}
\caption{Real part of the $s$-wave pion selfenergy for the pion at rest. Note that for asymmetric nuclear matter, $k_F$ of the proton is plotted on the abscissa and we always take $k_F^n=1.157\,k_F^p$. Fit results to pionic atom data from Refs. \cite{Nieves:1993ev,Stricker:1980vm,Tauscher:1974bx} are also plotted. Theoretical calculation of Ref. \cite{Garcia-Recio:1987ik} indicated as ''Garcia et al.'' (dashed line).}
\label{fig:Pi_overview}
\end{figure}

In Fig. \ref{fig:Pi_overview}, the real part of the $s$-wave pion selfenergy from the full model and from several approximations is plotted. 
\begin{figure}
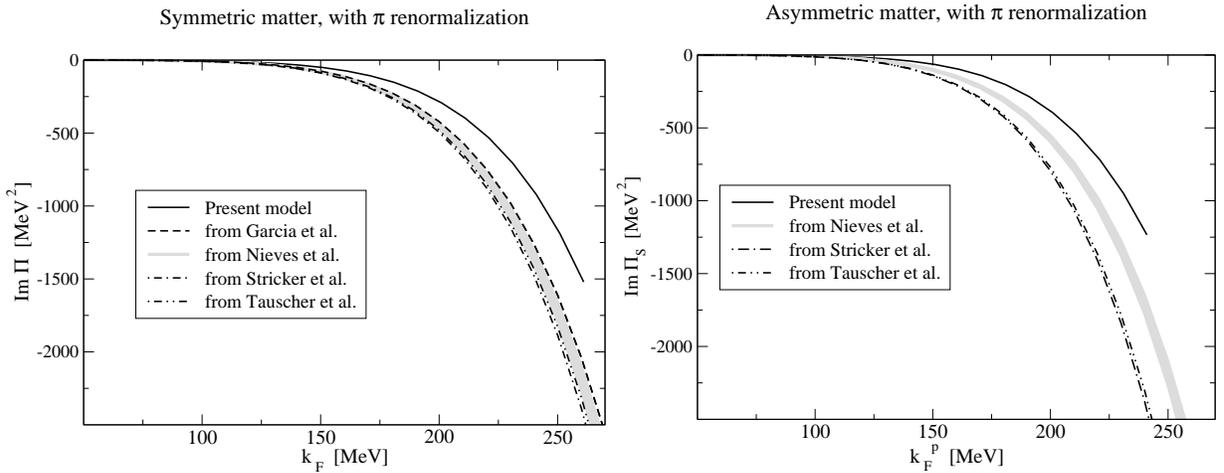

\includegraphics[width=8cm]{fig071_op.eps}
\includegraphics[width=8cm]{fig072_op.eps}
\caption{Imaginary part of the $s$-wave pion selfenergy for the pion at rest. Phenomenological fits as in Fig. \ref{fig:Pi_overview}.}
\label{fig:im_Pi_overview}
\end{figure}
The solid lines show the results for the model from Sec. \ref{sec:full_model} in symmetric and asymmetric nuclear matter, with and without the $p$-wave renormalization from Eq. (\ref{migdal}) of the pion propagator in the intermediate $\pi N$ loops. For the cases with asymmetric matter, the x-axis is given by $k_F^p$. The neutron Fermi momentum is then chosen to be $k_F^n=1.154 k_F^p$. This ratio corresponds to the ratio of neutron rich nuclei as ${}^{208}_{82}{\rm Pb}$ with $k_F^p=241$ MeV and $k_F^n=278$ MeV. The selfenergy in asymmetric nuclear matter is larger than in symmetric matter which can be easily understood from the large and positive term $(-4\pi)b_1(\rho_n-\rho_p)$ from Eq. (\ref{param_opt}). 

The effect of Pauli blocking in the intermediate loops of the $s$-wave rescattering (see Eq. (\ref{matterg})) can be taken into account by the Ericson approximation \cite{Ericson:1966fm,Garcia-Recio:1987ik} 
\be
\Delta b_0 (k_F)=-\frac{6k_F}{\pi\, m_\pi^2}\frac{m_N}{m_\pi+m_N} \left(\lambda_1^2+2\lambda_2^2\right).
\label{ericson_correction}
\ee
As pointed out in Ref. \cite{Garcia-Recio:1987ik} the quantities $\lambda_{1,2}$ are related to the vacuum isoscalar and isovector $b_0, b_1$ terms (generated from rescattering, not the elementary ones) for which we take from Ref. \cite{Doring:2004kt}, 
\begin{align}
b_{0,\;{\rm vac}}&=-0.0028\; m_\pi^{-1}{\hat =}-\frac{1}{1+\frac{m_\pi}{m_N}}\; \frac{2\lambda_1}{m_\pi},\non
b_{1,\;{\rm vac}}&=-0.0881\;m_\pi^{-1}{\hat =}-\frac{1}{1+\frac{m_\pi}{m_N}}\; \frac{2\lambda_2}{m_\pi}. 
\label{vacbo}
\end{align}
With these values and $B_0=0$ (no pion medium modification) one obtains from Eq. (\ref{param_opt}) the dotted curve in Fig. \ref{fig:Pi_overview} upper left panel. Adding the approximate medium change of $b_0$ from Eq. (\ref{ericson_correction}) according to $b_0=b_{0,\;{\rm vac}}+\Delta b_0$, the dashed curve is obtained. Thus, the $t\rho$ approximation is not sufficient, whereas the inclusion of $\Delta b_0$ leads to a good agreement with the rescattering model. This shows also that effects from Pauli blocking in more than one loop in the $\pi N$ rescattering of the $\pi N$ amplitude are small, because Eq. (\ref{ericson_correction}) corresponds to exactly one Pauli blocked loop in the rescattering series \cite{Garcia-Recio:1987ik}. 

Next, we compare to asymmetric nuclear matter but still without pion modification. This is displayed in Fig. \ref{fig:Pi_overview} upper right panel. Now, the isovector term contributes and we can derive a similar approximation as Eq. (\ref{ericson_correction}) for the $b_1$ renormalization in nuclear matter,
\be
\Delta b_1 (k_F)=-\frac{6k_F}{\pi m_\pi^2}\frac{m_N}{m_\pi+m_N} \left(2\lambda_1\lambda_2-\lambda_2^2\right)
\label{b1reno}
\ee 
by simply comparing the isospin structure of $\pi N$ scattering at one loop. The result from Eq. (\ref{param_opt}) using $b_0=b_{0,\;{\rm vac}}+\Delta b_0$ and $b_1=b_{1,\;{\rm vac}}$ is indicated as the dotted line. As the dashed line we plot the result from Eq. (\ref{param_opt}) using $b_0=b_{0,\;{\rm vac}}+\Delta b_0$ and $b_1=b_{1,\;{\rm vac}}+\Delta b_1$. Obviously, the correction from Eq. (\ref{b1reno}) is small. However, in Sec. \ref{sec:higher_order} we will find additional vertex corrections that will modify appreciably the isovector strength of $\pi N$ scattering.

When including the pion renormalization in the model according to Eqs. (\ref{matterg},\ref{spectral}) the real part of the $s$-wave pion selfenergy for symmetric and asymmetric nuclear matter decreases as shown in the two lower plots of Fig. \ref{fig:Pi_overview}. We can compare to Ref. \cite{Garcia-Recio:1987ik}. For this, we take the final values for $B_0$ from there, $B_0=0.032+i\,0.040 \;m_\pi^{-4}$. Note that this is only qualitative because we do not take the density dependence of $B_0$ from Ref. \cite{Garcia-Recio:1987ik} into account but use a mean value. The values from \cite{Garcia-Recio:1987ik} for $b_0$ and $b_1$ are $-0.013\,m_\pi^{-1}$ and $-0.092\,m_\pi^{-1}$, respectively. With these values and adding $\Delta b_0$ from Eq. (\ref{ericson_correction}) to $b_0$, the selfenergy is calculated according to Eq. (\ref{param_opt}) and plotted in Fig. \ref{fig:Pi_overview}, lower left panel for symmetric nuclear matter. In the same plot the $s$-wave selfenergy from fits to the bulk of pionic atom data from Refs. \cite{Tauscher:1974bx,Stricker:1980vm,Nieves:1993ev} with the values given in Tab. \ref{tab:bobo} is shown. Both the present model and results from Ref. \cite{Garcia-Recio:1987ik} are systematically below the phenomenological values. Neither the present model nor Ref. \cite{Garcia-Recio:1987ik} reach the required size for the real part of $\Pi_S$ and thus the problem of missing repulsion persists. 

The imaginary part of the pion $s$-wave selfenergy is displayed in Fig. \ref{fig:im_Pi_overview}. The result from Ref. \cite{Garcia-Recio:1987ik} (dashed line) agrees well with the phenomenological values from Refs. \cite{Tauscher:1974bx} and \cite{Stricker:1980vm} (gray band) whereas the present model shows a 30\% discrepancy. 

The differences between the results from Ref. \cite{Garcia-Recio:1987ik} and the present calculation (dashed vs. solid line for the symmetric matter case including the pion renormalization) should be attributed to a different input used in \cite{Garcia-Recio:1987ik}, such as form factors plus the fact that extra crossed terms of $\rho^2$ character (smaller than those incorporated here) were also evaluated in \cite{Garcia-Recio:1987ik}. The larger repulsion from \cite{Garcia-Recio:1987ik} can be partly explained by the large vacuum $|b_0|, \,|b_1|$ used there, whereas nowadays values for $b_0$, compatible with zero as in Eq. (\ref{vacbo}), are regarded as more realistic. 


\subsection{Self consistent treatment of the amplitude}
\label{sec:selfcon}
For the pion polarization in intermediate $\pi N$ loops, so far only the $p$-wave pion selfenergy has been taken into account. For the $s$-wave part we can include the selfenergy determined in the last section in a self consistent approach. For this, the $\pi^-$ selfenergy $\Pi_S$ from Eq. (\ref{sum_fermi}) is included in the pion propagator from Eq. (\ref{spectral}). Additionally, the selfenergy is resummed so that it can be included in the same way as the $p$-wave selfenergy $\Pi_{p,\,\pi}$ in the pion propagator,
\be
D_{\pi}=\frac{1}{(q^0)^2-{\bf q}^2-m_{\pi_i}^2-
\Pi_{p,\,\pi}(q^0, {\bf q};\rho_p, \rho_n)-\Pi_S(q^0=m_\pi, {\bf q}=0;\rho_p,\rho_n)}.
\label{self_cons}
\ee 
We have approximated here the energy and momentum dependence of $\Pi_S$ by the static case $(q^0=m_\pi, {\bf q}=0)$.
Solving for $\Pi_S$ by iteration one obtains the results in Tab. \ref{tab:iteration} for asymmetric matter. As in Sec. \ref{sec:numres} we set $k_F^n=1.157\,k_F^p$ and show the results for $k_F^p=213$ MeV and $k_F^p=241$ MeV which corresponds to densities of around $\rho_0/2$ and $\rho_0$. 
\linespread{1.2}
\begin{table}
\caption{Self consistent treatment of the $s$-wave selfenergy $\Pi_S(q^0=m_\pi, {\bf q}={\bf 0})$ in [MeV$^2$] for asymmetric matter. To the left, the case with $k_F^p=213$ MeV, to the right $k_F^p=241$ MeV. Three iteration steps are shown.}
\begin{tabular*}{0.6\textwidth}{@{\extracolsep{\fill}}lllll}
&Re$(\Pi_S)_{[213\,{\rm MeV}]}$&Im$(\Pi_S)_{[213\,{\rm MeV}]}$&Re$(\Pi_S)_{[241\,{\rm MeV}]}$&Im$(\Pi_S)_{[241\,{\rm MeV}]}$
\\ \hline \hline
Step 0&$2470.4$&$-570.8$&$3423.6$&$-1233.8$ \\
Step 1&$2503.9$&$-562.4$&$3491.3$&$-1207.4$ \\
Step 2&$2504.3$&$-562.1$&$3492.2$&$-1205.8$ 
\\ \hline
\end{tabular*}
\label{tab:iteration}
\end{table}
\linespread{1.0}
Three iteration steps are shown with step 0 being the selfenergy without iteration. Comparing the size of $\Pi_S$ from Figs. \ref{fig:Pi_overview} and \ref{fig:im_Pi_overview} with $m_\pi^2$ from the propagator, the result is expected to change only little. Indeed, the iteration converges rapidly and changes are small. 
At this point one can improve the calculation by evaluating the $s$-wave pion self energy not in the approximation $(q^0=m_\pi, {\bf q}={\bf 0})$ as in Eq. (\ref{self_cons}), but with the full $q^0,\,{\bf q}$ dependence: it is known, at least for the vacuum case, that the isoscalar $\pi N$ amplitude is small at threshold but then grows rapidly at finite scattering energies. Taking only the $q^0$-dependence --- the ${\bf q}$ dependence is small --- the self consistent calculation delivers indeed a larger change than before, of about 10 \% of additional repulsion at $\rho=\rho_0/2$.

\section{Higher order corrections of the isovector interaction}
\label{sec:higher_order}
In this section additional corrections are introduced that go beyond the medium modifications from Sec. \ref{sec:full_model}, namely medium corrections affecting the kernel of the Bethe-Salpeter equation itself. In our model the kernel is given by the Weinberg-Tomozawa isovector $\pi N\to\pi N$ transition and the NLO isoscalar $\pi N\to  \pi N$ transition. Considering vertex corrections of the rescattering is advantageous because it allows to include higher order corrections to the Ericson-Ericson rescattering piece, that is a large source of isoscalar strength. The corresponding $s$-wave pion selfenergy diagrams appear at higher orders in density that are difficult to access through a systematic expansion of the selfenergy (see, e.g., Ref. \cite{Kaiser:2001bx}). In this section we will consider the renormalization of the Weinberg-Tomozawa interaction through vertex corrections. In Sec. \ref{sec:renoiso} similar changes to the NLO isoscalar piece will be applied. From now on only symmetric nuclear matter will be considered.

\subsection{Tadpoles and off-shell contributions}
\label{sec:renotomo}
In the vacuum the vertex renormalizations can be partly absorbed in the coupling constant $f_\pi$. In the nuclear medium, these diagrams should be explicitly taken into account.
\begin{figure}
\includegraphics[width=0.7\textwidth]{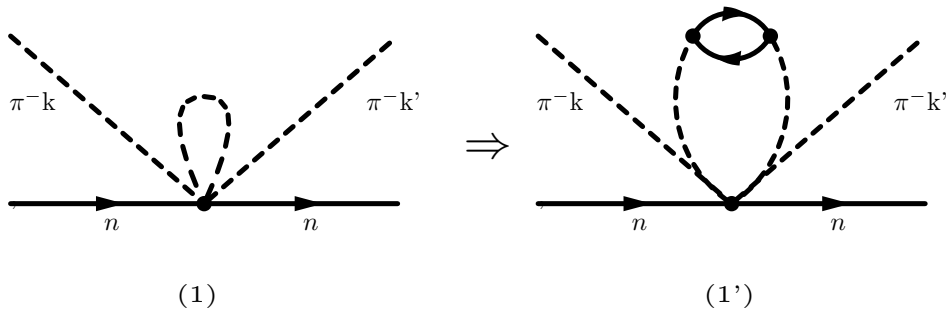}
\caption{The vertex tadpole at $1/f_{\pi}^4$ (diagram 1) and the corresponding medium diagram at $1/f_{\pi}^6$ (diagram 1').}
\label{fig:first_vertex}
\end{figure}
Fig. \ref{fig:first_vertex} (1) shows a tadpole diagram that involves a four-pion nucleon vertex. In the free case, this term is accounted for implicitly through a renormalization of the lowest order Weinberg-Tomozawa term. However, in the medium the virtual pion can be polarized by exciting $ph$ or $\Delta h$ excitations and this leads to diagram (1') of Fig. \ref{fig:first_vertex}. The difference between these two terms should be considered a genuine many body correction. 

A diagram with the same geometry but within a linear $\sigma$ model has been also proposed in \ccc{Chanfray:2003js}. The $4\pi2N$ vertex in diagram (1) of Fig. \ref{fig:first_vertex} is obtained from the LO chiral Lagrangian with two baryons,
\be
{\cal L}^{(2)}_{\pi N}=i\;{\rm Tr}\left[\bar{B}\gamma^\mu\left[\Gamma_\mu, B\right]\right]
\label{4pil1}
\ee
with $\Gamma_\mu$ expanded up to four meson fields,
\be
\Gamma_\mu=\frac{1}{32f_\pi^4}\left[\frac{1}{3}\;\partial_\mu\Phi\Phi^3-\Phi\partial_\mu\Phi\Phi^2+\Phi^2\partial_\mu\Phi\Phi-\frac{1}{3}\; \Phi^3\partial_\mu\Phi\right]
\label{4pil2}
\ee
where $\Phi$ is the standard $SU(2)$ representation of the pion field, $\Phi_{11}=1/\sqrt{2}\pi^0$, $\Phi_{12}=\pi^+$, $\Phi_{21}=\pi^-$, $\Phi_{22}=-1/\sqrt{2}\pi^0$.
For the process $\pi^-n\to\pi^- n$ where the external pions have on-shell momenta $k, k'$ the diagrams (1) and (1') are given by
\be
V^{(1),\,(1')}_{\pi^-n\to\pi^-n}=-\frac{5}{48}\frac{1}{f_\pi^4}\;(k^0+k'^0)\sqrt{\frac{E_i+M_i}{2M_i}}\sqrt{\frac{E_j+M_j}{2M_j}}\;i\int\frac{d^4 p}{(2\pi)^4}\; D_{(1),\,(1')}(p)
\label{V_diagrams}
\ee
which can be approximated by $k^0+k'^0=2\sqrt{s}-M_i-M_j$ with $M_i, M_j$, $E_i, E_j$ the masses and energies of the incoming and outgoing nucleons $i$ and $j$. In Eq. (\ref{V_diagrams}) we have made the same $s$-wave projection as for the ordinary $\pi N\to\pi N$ amplitude \cite{Doring:2004kt,Inoue:2001ip}.
The meson propagators for diagram (1) and (1') are given by
\be
D_{(1)}=\frac{1}{p^2-m_\pi^2+i\epsilon},\quad D_{(1')}=\int\limits_0^\infty d\omega\;\frac{2\omega S_\pi(\omega,{\bf p},\rho)}{(p^0)^2-\omega^2+i\epsilon}
\label{dprop}
\ee
where $S_\pi$ is the pion in-medium spectral function from Eq. (\ref{spectral}). The contribution of the vertex correction can then be written as a correction to the kernel $V\to V+\delta V$ of the Bethe-Salpeter Eq. (\ref{bse}) where $\delta V=V^{(1')}-V^{(1)}$. This is because $D_{(1')}$ contains also $D_{(1)}$ and the vacuum diagram has to be subtracted explicitly.

One can see from Eq. (\ref{V_diagrams}) that $\delta V$ has explicitly order $1/f_\pi^4$. However, in Eq. (\ref{dprop}) $D_{(1')}-D_{(1)}$ is of order $1/f_\pi^2$ (and higher from $ph$, $\Delta h$ iterations in the spectral function $S_\pi$) since the $ph$ excitation $p$-wave pion selfenergy is of order  $1/f_\pi^2$. Thus, the correction $\delta V$ is of order $1/f_\pi^6$ and higher. 

By looking at other transitions such as $\pi^-p\to\pi^-p$ or $\pi^-p\to\pi^0 n$, we observe that the vertex contributions from Eqs. (\ref{4pil1}, \ref{4pil2}) are of isovector nature. This means that one can absorb the vertex correction as a common factor in the definition of $f_\pi$, as it appears in the isovector amplitude, resulting in an in-medium renormalized $f_{\pi,\;{\rm med}}^2$ for the isovector term,
\be
\frac{b_1^*(\rho)}{b_{1\,{\rm free}}}\equiv
\frac{f_\pi^2}{f_{\pi,\;{\rm med}}^2(\rho)}=1+\frac{r}{f_\pi^2}\int \limits_0^\infty \frac{dp\,p^2}{2\pi^2}\Big(-\frac{1}{2\eta}+\int\limits_0^\infty d\omega\; S_\pi(\omega,p,\rho)\Big)
\label{multif}
\ee
with $r=-5/12$ and $\eta^2=p^2+m_\pi^2$. 
In Eq. (\ref{multif}), $b_1^*(\rho)$ and $b_{1\,{\rm free}}$ are the density dependent isovector term and vacuum isovector term, respectively.
Although the diagrams (1) and (1') are linearly divergent, their difference, which gives the medium correction, is not; thus, the $p$-integration in Eq. (\ref{multif}) is well defined. Note that casting the vertex correction as a correction to the coupling $f_\pi$ in Eq. (\ref{multif}) is just for convenience. E.g. for the $\pi NN$ $p$-wave coupling, where $f_\pi$ also appears, such a procedure does not apply. Hence, the warning here is that one should be careful not to talk about a universal renormalization of $f_\pi$. 
It is worth noting that for $\rho=0$, the vacuum spectral function is given by
\be
S_\pi(\omega,p,\rho)\to \frac{1}{2\eta}\;\delta(\omega-\eta).
\label{svac}
\ee
We observe that the integral in Eq. (\ref{multif}) indeed vanishes for $\rho=0$.

\begin{figure}
\includegraphics[width=0.7\textwidth]{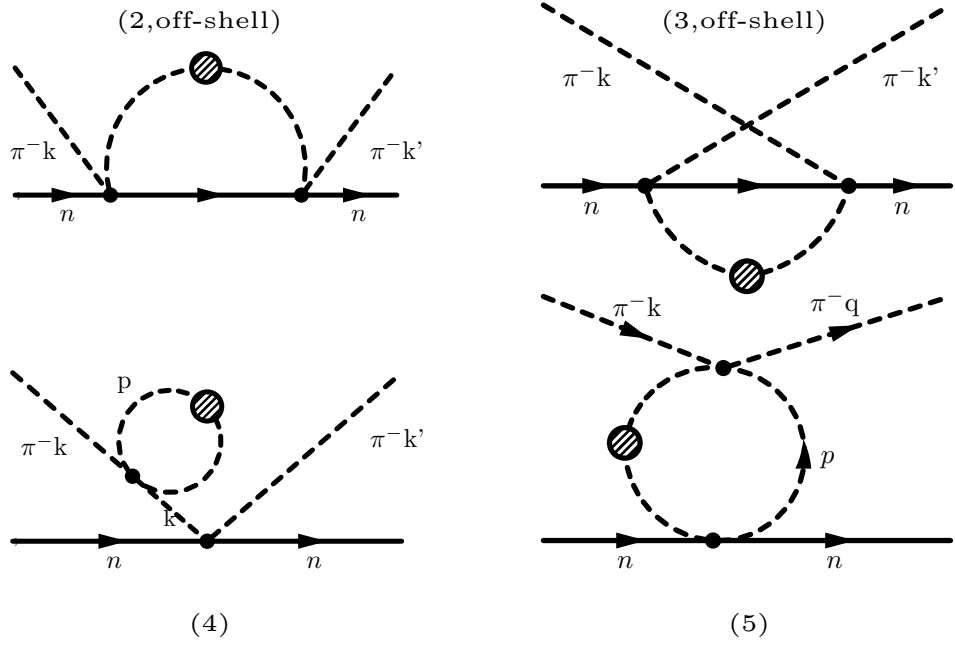}
\caption{Additional medium renormalizations at $1/f_{\pi}^6$ and higher. Off-shell parts of direct and crossed term are indicated in diagrams (2) and (3). Renormalization of the pion propagator in (4) and additional vertex correction with a loop in the $t$-channel, diagram (5). The shaded circles indicate resummed insertions of $ph$, $\Delta h$ pion $p$-wave selfenergies in the pion propagator, including also $NN$, $N\Delta$, $\Delta\Delta$ short range correlations (SRC).}
\label{fig:more_higher}
\end{figure}
Next, we turn to another kind of contribution.
In the on-shell reduction scheme of the $\pi N$ amplitude from Ref. \cite{Doring:2004kt} the on-shell and off-shell part of the $\pi N$ loop is separated and it can be shown that the off-shell part can be absorbed in the coupling of the $\pi N$ interaction \cite{Oset:1997it}. However, in the nuclear medium, this is no longer the case and one has to take the off-shell part explicitly into account. In the free case, the off-shell part in the vertices of the rescattering diagram (2) and the crossed diagram (3) in Fig. \ref{fig:more_higher} cancel the intermediate nucleon propagator in the heavy baryon limit, leading to a diagram with the same structure as (1) in Fig. \ref{fig:first_vertex}. As an example we consider $\pi^-n\to\pi^-n$ scattering via a $\pi^- n$ loop as shown in diagram (2) in Fig. \ref{fig:more_higher}. The amplitude is then, with $k=k'$ ($p$) the momentum of the external $\pi^-$ (external neutron) and $q$ the momentum of the $\pi^-$ in the loop,
\be
V^{(2)}_{\pi^-n\to\pi^-n}=i\int\frac{d^4q}{(2\pi)^4}\;\frac{M_N}{E({\bf q})}\left(\frac{k^0+q^0}{4f_\pi^2}\right)^2\frac{1}{k^0+p^0-q^0-E({\bf q})+i\epsilon}\;\frac{1}{q^2-m_\pi^2+i\epsilon}.
\label{exaonshell}
\ee
Using in the heavy baryon approach $p^0-E({\bf q})\sim 0$ and expanding the numerator as $\left(2k^0+q^0-k^0\right)^2=4(k^0)^2+4k^0(q^0-k^0)+(q^0-k^0)^2$, the on-shell part is given by the $4(k^0)^2$-term. For the other terms, the baryon propagator is canceled and Eq. (\ref{exaonshell}) reads 
\be
V^{(2)}_{\pi^-n\to\pi^-n}\approx V_{\rm on} G_{\pi N} V_{\rm on}+(2k^0)\left(\frac{-3i}{32f_\pi^4}\right)\int\frac{d^4q}{(2\pi)^4}\;\frac{1}{q^2-m_\pi^2+i\epsilon}
\label{deltav2}
\ee
with $V_{\rm on}$ the usual on-shell transition $\pi^-n\to\pi^-n$ and $G_{\pi N}$ the $\pi^-n$ loop function. In Sec. \ref{sec:full_model} medium corrections have been applied to the first term in Eq. (\ref{deltav2}), the on-shell one-loop rescattering. The second term is the product of the usual $\pi N$ on-shell amplitude times a pion tadpole and has, thus, the same structure as diagram (1) of Fig. \ref{fig:first_vertex}. The remaining pion tadpole is dressed in the way it is done for diagram (1') of Fig. \ref{fig:first_vertex} and the vacuum tadpole is subtracted; the result can again be expressed in a renormalization of $f_\pi$ in Eq. (\ref{multif}), this time with $r=-3/8$. 
The crossed term in $\pi^-n\to\pi^-n$ scattering via one loop is displayed in Fig. \ref{fig:more_higher}, (3). Note that the intermediate states are in this case $\pi^+n$ and $\pi^0 p$. Evaluating the off-shell parts as before, again the structure of tadpole and on-shell scattering of diagram (1') is obtained. Summing both off-shell parts from diagrams (2) and (3) the result can be cast in a modification of $f_\pi$ as in Eq. (\ref{multif}) with $r=+3/4$. The calculation is repeated for the other coupled channels $\pi^-p\to\pi^-p$, $\pi^-p\to\pi^0n$, and $\pi^0n\to\pi^0n$ and it is interesting to note that the off-shell parts of the one-loop amplitude have pure isovector character. This is in contrast to the on-shell one-loop amplitude with two pure isovector scatterings that results in a mixture of isovector and isoscalar contributions.

In addition we have to consider structures as in Fig. \ref{fig:more_higher} (4), (5) at the same order in $f_\pi$ and density. For the tadpole pion selfenergy in diagram (4) of Fig. \ref{fig:more_higher} we consider the process $\pi^-n\to\pi^-n$ with the external pions at momentum $k$ and the  $\pi\pi$ vertex given by the LO chiral Lagrangian.
The vacuum $\pi^-$ selfenergy of this external pion line consists in charged and neutral pion loops and can be written as
\be
(-i\Pi)=\frac{1}{6f_\pi^2}\int\frac{d^4 p}{(2\pi)^4}\;\left(4(p^2-m_\pi^2)+4(k^2-m_\pi^2)+5m_\pi^2\right)\;\frac{1}{p^2-m_\pi^2}.
\label{pi4}
\ee
We have written the momentum structure from the $\pi\pi$ vertex in a form where it becomes visible that the first and third term contribute to the pion wave function renormalization. These terms can be incorporated in the free pion mass. Continuing with the free case, the pion tadpole (4) can appear attached to an intermediate pion in the rescattering scheme (see Fig. \ref{fig:bse}). In this case, one of the two intermediate pion propagators of momentum $k$ cancels the $k^2-m_\pi^2$ structure of the second term in Eq. (\ref{pi4}). As a consequence, a tadpole attached to the $\pi N$-vertex results, with the structure of diagram (1) in Fig. \ref{fig:first_vertex}. The medium corrections arise then from the dressing of the pion as displayed in diagram (1').

However, the pion tadpole (4) from Fig. \ref{fig:more_higher} can also appear in an external pion line of the rescattering displayed in Fig. \ref{fig:bse}. 
In this case, the first and third term of Eq. (\ref{pi4}) contribute to the external pion wave function renormalization in the medium. In other words, this is a reducible diagram, because two pieces are separated by a pion propagator. In the search for pion selfenergy terms we must only look for irreducible diagrams. However, the second term in the bracket is special because it exactly cancels the pion propagator $(k^2-m_\pi^2)^{-1}$ leading to a genuine irreducible diagram, that must be taken into account and is of the tadpole type of Fig. \ref{fig:first_vertex} (1'). 

Inserting the pion tadpole in this way in internal as well as external pion lines,
the corresponding $\delta V^{(4)}$ from Fig. \ref{fig:more_higher} (4) is given by 
\be
\delta V^{(4)}=\frac{1}{k^2-m_\pi^2}\left(\frac{2k^0}{4f_\pi^2}\right)\left(k^2-m_\pi^2\right)\frac{2}{3f_\pi^2}\,i\int\frac{d^4p}{(2\pi)^4}\;
(D_{(1')}-D_{(1)})
\label{delta4}
\ee
which, by analogy to the terms calculated before, can be recast into a renormalization of $f_\pi$ (for the purpose of the isovector term) given in Eq. (\ref{multif}) with $r=2/3$. In this case the isovector character is obvious as the pion selfenergy is the same for all charge states of the pion.

Note that there should be a symmetry factor of 2 as one can insert the pion tadpole also at the other pion line in diagram (4). However, if the pion selfenergy is inserted in an intermediate $\pi N$ loop of the rescattering series, this symmetry factor is not present --- each intermediate pion has only one pion selfenergy insertion. 
Note that for the contribution from inserting the pion tadpole in an external pion line of the rescattering scheme of Fig. \ref{fig:bse} there is a factor $1/2$ to be taken into account in the wave function renormalization when considering the adiabatic introduction of the interaction \cite{mandl}. Considering this, it is easy to see that Eq. (\ref{delta4}) takes already correctly into account all multiplicity factors.

\subsection{Loop corrections in the $t$-channel}
\label{sec:dia5isovector}
For the vertex correction (5) in Fig. \ref{fig:more_higher} we consider the process  $\pi^-p\to\pi^-p$. The loop of the vertex correction is charged, because a neutral pion in the loop can not couple to the Weinberg-Tomozawa term. 
The diagram will be evaluated for forward scattering $k=q$ which simplifies the calculation --- this kind of approximation will be made several times in the following and is discussed below. Then, the vacuum amplitude for $\pi^-p\to\pi^-p$ is given by
\be
(-it)^{(5)}=-\frac{1}{6\,f_\pi^4}\int\frac{d^4p}{(2\pi)^4}\,D^2(p)\,p^0\,\left(p^2+6\,pq+q^2-2m_\pi^2\right).
\label{eqn5}
\ee
The pion propagators $D(p)$ are given by Eq. (\ref{dprop}). The term $p^0$ comes from the Weinberg-Tomozawa vertex and the momentum structure in the brackets is from the $\pi\pi$ vertex with momentum $q$ for the external pions. By symmetric integration, the only non-vanishing structure is given by the combination $6(p^0)^2q^0$. As an explicit calculation shows, this one-loop correction is again of isovector type. 

For the vertex correction in the nuclear medium, one of the propagators $D$ in Eq. (\ref{eqn5}) is dressed according to $D^2\to D_{(1)}\,D_{(1')}$ with the definitions from Eq. (\ref{dprop}).
Then, a factor of two is supplied in order to account for the two possibilities to insert the medium dressing in either of the intermediate propagators. We have checked this approximation by performing the full calculation with medium dressings in both propagators which leads to negligible higher order corrections. 

A straight evaluation of diagram (5) along the lines  used for the other diagrams gives a contribution of the same size as the others. However, it is easy to see that this would be a gross overestimation. The contribution from diagram (5) can be disregarded as shown in the following. In the first place, the Weinberg-Tomozawa term, extrapolated to the high energies $p^0$ involved in the loop, is a gross overestimation of the actual isovector $\pi N$ amplitude. In order to quantify this, the isovector amplitude from the unitary coupled channel model has been evaluated, with the set of parameters given by the right column of Tab. \ref{tab:parms}. This provides a realistic amplitude up to the region of the $N^*(1535)$. Second, the (momentum dependent) ratio of the isovector amplitude of this model to the amplitude from the Weinberg-Tomozawa interaction has been determined. This ratio can be well approximated by a scale factor
\be
F_I(p)=\frac{\Lambda_1}{\sqrt{\Lambda_2^2+p^2}}
\label{addff}
\ee
with $p$ the c.m. $\pi N$ three-momentum, $\Lambda_1=225$ MeV and $\Lambda_2=200$ MeV. 
Including this scale factor, the medium contribution from diagram (5) is given by
\be
\frac{\delta V^{(5)}}{V_{\rm WT}}=\frac{2}{f_\pi^2}\int\limits_0^\infty\frac{dp\;p^2}{2\pi^2}\;F_I(p)\;\left(-\frac{1}{2\eta}+\int\limits_0^\infty d\omega\;\frac{2\omega}{\eta+\omega}\;S_\pi(\omega,p,\rho)\right)
\label{newnew}
\ee
where we have additionally divided by the Weinberg-Tomozawa term $V_{\rm WT}$.


We have calculated diagram (5) in the forward scattering limit, and furthermore assuming the external pions at rest. This overestimates the contribution as it is easy to see: consider diagram (5) where one external pion line corresponds to the pion at rest in the nuclear medium whereas the other pion line is in a rescattering loop. Then, one pion is at momentum $(m_\pi,0)$ whereas the pion in the loop can take high values of $q^0,{\bf q}$. For a typical loop momentum of $|{\bf q}|=1$ GeV this momentum mismatch leads to a reduction of a factor of around two for the diagram. Together with the scale factor from Eq. (\ref{addff}), this additional reduction renders the contribution from diagram (5) small, smaller than the theoretical uncertainties which will be summarized in Sec. \ref{isovec_reno}. Thus, diagram (5) is neglected.

\begin{figure}
\includegraphics[width=5cm]{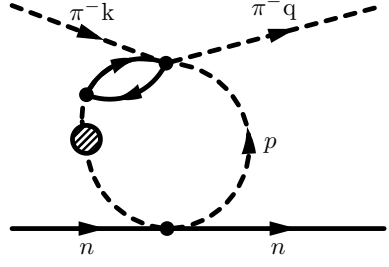}
\caption{Loop in the $t$-channel with a different $ph$-insertion.}
\label{fig:twotad22}
\end{figure}

In Fig. \ref{fig:twotad22}
another in-medium diagram is shown with a $ph$ or $\Delta h$ directly coupling to three pions. The $3\pi NN$-interaction is obtained from an expansion of the part with $D,\,F$ of the LO chiral $\pi N$ Lagrangian up to three pion fields \cite{Oset:2000gn,Kaskulov:2005kr}. This interaction provides terms of the form $\boldsymbol{\sigma}\cdot {\bf p}$ and $\boldsymbol{\sigma}\cdot {\bf q}$ with the three-momenta ${\bf p, q}$ of the loop and the external pions, respectively. As the leading term of the Weinberg-Tomozawa interaction is of the form $p^0$, the diagram vanishes by symmetric integration.

\begin{figure}
\includegraphics[width=5cm]{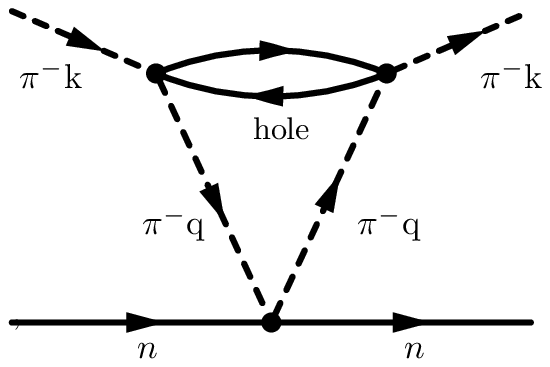}\hspace*{1cm}
\includegraphics[width=5cm]{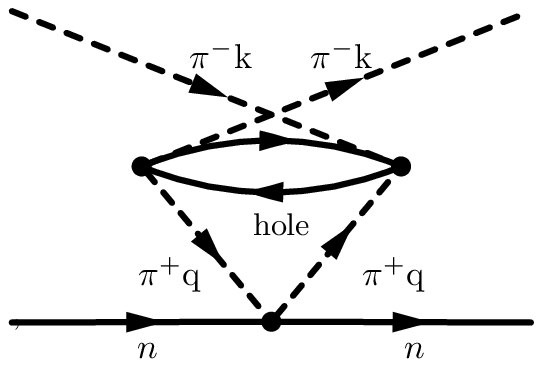}
\caption{Additional vertex corrections. The two diagrams come with a relative minus sign due to the isovector character of the $\pi N$ $s$-wave interaction.}
\label{fig:bull}
\end{figure}

In the following we would like to discuss another type of loop corrections in the $t$-channel, the diagrams of Fig. \ref{fig:bull}.
The sum of the two diagrams involves the contribution $U_d(q^0+k^0)-U_d(q^0-k^0)$, with $U_d$ the Lindhard function for only forward going bubbles. Terms involving this combination are found very small in Appendix B of \cite{Cabrera:2000dx} and we do not consider them. 

We do not consider selfenergy insertions in the nucleon lines. The reason is that summing over occupied states in Eq. (\ref{sum_fermi}) corresponds to a $ph$ excitation; a local selfenergy in the particle and the hole lines cancels in the $ph$ propagator. We will come back to this question in Sec. \ref{sec:ordering}.

\subsection{Vertex corrections from $\boldsymbol{\pi NN}$ and $\boldsymbol{\pi N\Delta}$ related terms}
\label{sec:NLO}
\begin{figure}
\includegraphics[width=0.3\textwidth]{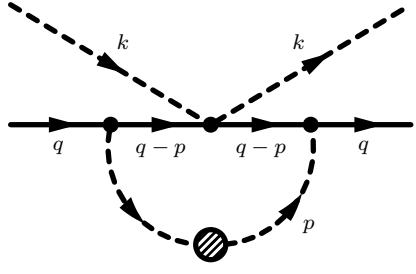}\hspace*{0.7cm}
\caption{Additional vertex correction. Dressing the pion and introducing Pauli blocking for the intermediate nucleons gives a density dependent correction of the isovector amplitude.}
\label{fig:another}
\end{figure}

Another renormalization of the isovector amplitude is shown in Fig. \ref{fig:another}.
The diagram exhibits two $p$-wave $\pi NN$ vertices; the same, important, short-range correlations between $ph$ and $\Delta h$, that are included in the dressed pion propagator (see Eq. (\ref{migdal})), should also be taken into account between a $p$-wave vertex of the diagram and the adjoint $ph$ or $\Delta h$ insertion in the pion propagator. The inclusion of these short range correlations (SRC) is most easily achieved by decomposing the pion selfenergy in a longitudinal and a transversal part $V_l$ and $V_t$. The matter part of the diagram after subtracting the vacuum loop, divided by the tree level Weinberg-Tomozawa term $V_{\rm WT}$, is given by
\be
\frac{\delta V^{(\ref{fig:another})}_{\rm SRC}}{V_{\rm WT}}&=&-i\int\frac{d^4 p}{(2\pi)^4}\left(\frac{M_N}{E_N({\bf p})}\;\frac{1}{M_N-p^0-E_N({\bf p})+i\epsilon}\right)^2
\Bigg(-\theta(k_F-|{\bf p}|)\left[(V_l(p^0,{\bf p})-V_t(p^0,{\bf p})\right]
\non
&+&
\theta(|{\bf p}|-k_F)\left[\frac{V_l(p^0,{\bf p})}{1-U(p^0,{\bf p})V_l(p^0,{\bf p})}+
\frac{2\,V_t(p^0,{\bf p})}{1-U(p^0,{\bf p})V_t(p^0,{\bf p})}-V_l(p^0,{\bf p})-2\,V_t(p^0,{\bf p})\right]\Bigg)
\label{example_SRC}
\ee
where 
\be
V_l(p^0,{\bf p})=\left(\frac{f_{\pi NN}}{m_\pi}\right)^2 F^2(|{\bf p}|)\left(\frac{{\bf p}^2}{(p^0)^2-{\bf p}^2-m_\pi^2+i\epsilon}+g'\right),\quad
V_t(p^0,{\bf p})=\left(\frac{f_{\pi NN}}{m_\pi}\right)^2 F^2(|{\bf p}|)g'
\label{selftranslong}
\ee
and $U$ is the sum of $ph$ and $\Delta h$ Lindhard functions, $U=U_N+(f_{\pi N\Delta}/f_{\pi NN})^2\,U_\Delta$. The monopole form factor $F$ and the Migdal parameter $g'$ have been already defined in Sec. \ref{sec:pwaveself}. The term with $\theta(k_F-p)$ accounts for the small correction from Pauli blocking of the intermediate nucleons without any modification of the pion, whereas the term with $\theta(p-k_F)$ comes from diagrams with pion polarization through $ph$ and $\Delta h$ insertions.

The $p^0$-integration is performed numerically. There is one technical complication resulting from the non-analyticity of the $\Delta$-width in the $\Delta h$ Lindhard function (step function $\Theta(\sqrt{s}-M_N-m_\pi)$; see Appendix of \cite{Oset:1989ey}). This leads to unphysical imaginary parts in $\delta V$ from the $p^0$-integration; the $\Delta$-width is, thus, set to zero for this diagram. The additional short-range correlations reduce the contribution from the diagram strongly. This is in agreement with findings from Ref. \cite{Kaskulov:2005kr} in the study of similar in-medium corrections for the isoscalar $NN$ interaction. 

\begin{figure}
\includegraphics[width=0.3\textwidth]{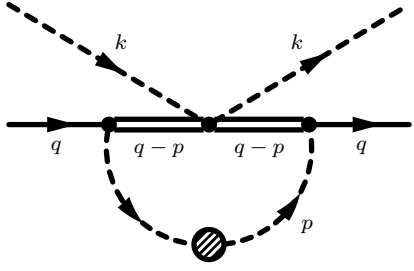}
\caption{The $\Delta$ as intermediate baryon in $\pi^-p\to\pi^-p$ scattering. The $\pi\Delta$-vertex is in $s$-wave and taken from \cite{Sarkar:2004jh}.}
\label{firstdelta}
\end{figure}

The intermediate nucleons in the diagram of Fig. \ref{fig:another} can also be excited. Close to threshold, even if it is off-shell, the $\Delta(1232)$ is important as we will see. The corresponding vertex correction is shown in Fig. \ref{firstdelta}.
For the $\pi\Delta\to\pi\Delta$ interaction we take the Weinberg-Tomozawa interaction of isovector type from Ref. \cite{Jenkins:1991es} in the $s$-wave approximation of Ref. \cite{Sarkar:2004jh}. 

As in case of the corresponding diagram with nucleons discussed above, the introduction of additional SRC for the two $\pi N\Delta$ vertices is important; using the same projection technique as above, the result reads
\be
\frac{\delta V^{(\ref{firstdelta})}_{\rm SRC}}{V_{\rm WT}}&=&i\;\frac{20}{9}\left(\frac{f_{\pi N\Delta}}{f_{\pi NN}}\right)^2\int\frac{d^4 p}{(2\pi)^4}\left(\frac{1}{M_N-p^0-E_\Delta+i\epsilon}\right)^2
\non
&\times&
\left[\frac{V_l(p^0,{\bf p})}{1-U(p^0,{\bf p})V_l(p^0,{\bf p})}+
\frac{2\,V_t(p^0,{\bf p})}{1-U(p^0,{\bf p})V_t(p^0,{\bf p})}-V_l(p^0,{\bf p})-2\,V_t(p^0,{\bf p})\right]
\label{add_short_range_both}
\ee
with $V_t$, $V_l$ from Eq. (\ref{selftranslong}), $M_N(M_\Delta)$ is the nucleon ($\Delta$) mass, $F$ is the monopole form factor for the off-shell pions at the $\pi N\Delta$ vertices, and $f_{\pi N\Delta}=2.13$ is the strong coupling of $\Delta$ to $\pi N$. An explicit evaluation of different charge states shows that the corrections from the diagrams of Fig. \ref{fig:another} and \ref{firstdelta} are of pure isovector nature. 

As before, the additional short-range correlations reduce the contribution from this diagram. The ratio $\delta V^{(\ref{firstdelta})}_{\rm SRC}/ V_{\rm WT}$ can reach values up to 0.4 at $\rho\sim\rho_0$ (and still 20 \% at $\rho\sim\rho_0/2$). A slight modification of Eq. (\ref{add_short_range_both}) comes from the fact that the $\pi N\Delta$ vertex, ${\bf S}\cdot {\bf p}$, implicitly used in Eq. (\ref{add_short_range_both}), is defined in the $\Delta$ rest frame; one has to boost the pion momentum to this frame. This leads to a reduction of the contribution by a factor of 0.68; yet the correction is large. 

Contrary to the case of the $\pi N$ scattering where we shall be able to get also many body vertex corrections to the isoscalar part from next to leading order terms (see Sec. \ref{sec:renoiso}), we have no control over the isoscalar $\pi\Delta$ interaction which accompanies the isovector one. It is, thus, inconsistent to consider only the isovector $\pi\Delta$ interaction as done here to see the effects on the $\pi N$ isoscalar part through rescattering. Thus, we shall not include this corrections when evaluating the isoscalar part coming from $\pi N$ rescattering and shall bear in mind that we have uncertainties from this source in the $\pi A$ isoscalar optical potential.

Additionally, there is a momentum mismatch if one of the external pion lines of Fig. \ref{firstdelta} corresponds to an external pion at rest whereas the other pion line is inside a rescattering loop. This mismatch affects the intermediate $\Delta$ propagators; for a typical loop momentum of ${\bf q}=1$ GeV, the diagram is reduced by another factor of about 2.5. Altogether, we attribute to this source an increase of $b_1$ from 10 to 20 \% at $\rho=\rho_0$, accepting this band as theoretical uncertainty.

\subsection{Triangle diagrams}
\label{sec:triangle_delta}
\begin{figure}
\includegraphics[width=1\textwidth]{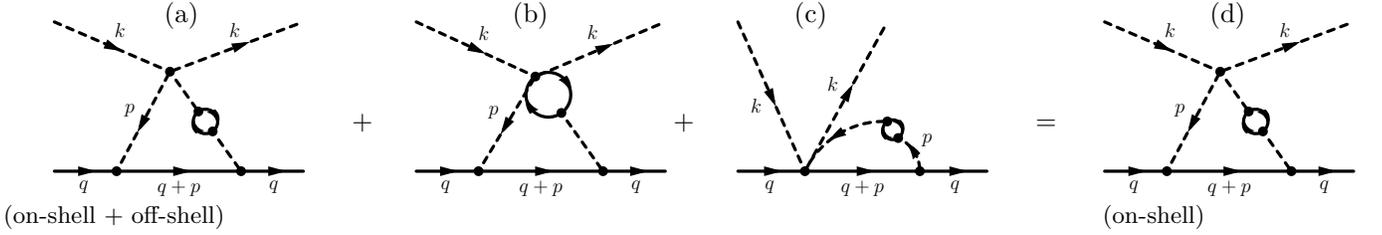}
\caption{Triangle diagrams. The labels ''off-shell'' and ''on-shell'' refer to the $\pi\pi$-vertex. Diagram (b) is complemented with a diagram with the $ph$-insertion in the other internal pion line. Diagram (c) is complemented with a diagram that has the loop on the other side of the $\pi N$ vertex.}
\label{fig:yet_another}
\end{figure}

There is another family of diagrams displayed in Fig. \ref{fig:yet_another}. As indicated in the figure, it is enough to calculate the diagram on the right hand side with the $\pi\pi$ vertex taken at its on-shell value in the sense that, whenever $p^2$ appears in the $\pi\pi$ amplitude, it has to be replaced by $m_\pi^2$. This is equivalent to calculating the same diagram on-shell plus off-shell, plus the set of other diagrams displayed on the left hand side of Fig. \ref{fig:yet_another}. This has has been shown in \cite{Oset:2000gn,Kaskulov:2005kr} and has been verified again in the present study for the limit of zero momentum exchange which is also the limit taken here for all vertex corrections. Such cancellations were first shown by \cite{Chanfray:1999nn,Chiang:1997di} in the problem of $\pi\pi$ scattering in the nuclear medium.

For the calculation, we consider first the reaction $\pi^-p\to\pi^-p$. For this configuration of external particles we can have charged or neutral pions for the loop lines. Summing both possibilities and inserting a factor of two from inserting the medium correction in either internal pion line, the medium amplitude takes the form 
\be
T^{(\ref{fig:yet_another},\, d)}&=&\frac{2i(D+F)^2}{3f_\pi^4}\int\frac{d^4p}{(2\pi)^4}\;F^2(|{\bf p}|)\,{\bf p}^2\;\frac{M_N}{E(p)}\non
&\times&
\int\limits_0^\infty d\omega\;\frac{2\omega\,S_\pi(\omega,p,\rho)}{(p^0)^2-\omega^2+i\epsilon}\;\frac{1}{(p^0)^2-\eta^2+i\epsilon}\;\frac{1}{p^0+M_N-E(p)+i\epsilon}\; \left(3p^0k^0+\frac{3}{4}\,m_\pi^2\right)
\label{ampln1}
\ee
where the momentum of the external (internal) pions is $k$ $(p)$, respectively. Again, we take the limit of forward scattering and, moreover, that the external pions and nucleons are at rest. In Eq. (\ref{ampln1}),  $E(p)=\sqrt{M_N^2+|{\bf p}|^2}$ is the nucleon energy. There is also a form factor $F(|{\bf p}|)$ for the off-shell pions in the $\pi NN$ vertex and the factor $M_N/E(p)$ from the non-relativistic reduction of the nucleon propagator.

A straightforward calculation reveals that the term $3p^0k^0$ is of isovector nature, whereas the contribution with $3/4 m_\pi^4$ is isoscalar. In the heavy baryon approximation we can neglect $M_N-E(p)$ in the baryon propagator and in this limit the isoscalar term cancels due to symmetric integration. We are left with a purely isovector contribution in which the $p^0$ from the numerator cancels the baryon propagator in the heavy baryon limit; the correction is given by
\be
\delta V^{(\ref{fig:yet_another},\, d)}&=&-(2k^0)\,\frac{(D+F)^2}{f_\pi^4}\int\limits_0^\infty\frac{dp\;p^2}{2\pi^2}\;p^2\;F^2(p)\;\frac{M_N}{E(p)}\frac{1}{\eta}\left(-\frac{\theta(k_F-p)}{8\eta^2}+\theta(p-k_F)\left[-\frac{1}{4\eta^2}+\int\limits_0^\infty d\omega\;\frac{S_\pi(\omega,p,\rho)}{\eta+\omega}\right]\right).\non
\label{correctyet}
\ee
The term with $\theta(k_F-p)$ accounts for the small medium correction from Pauli blocking of the intermediate nucleon but without any modification of the pion, whereas the term with $\theta(p-k_F)$ contains all diagrams with pion polarization.

\begin{figure}
\includegraphics[width=1\textwidth]{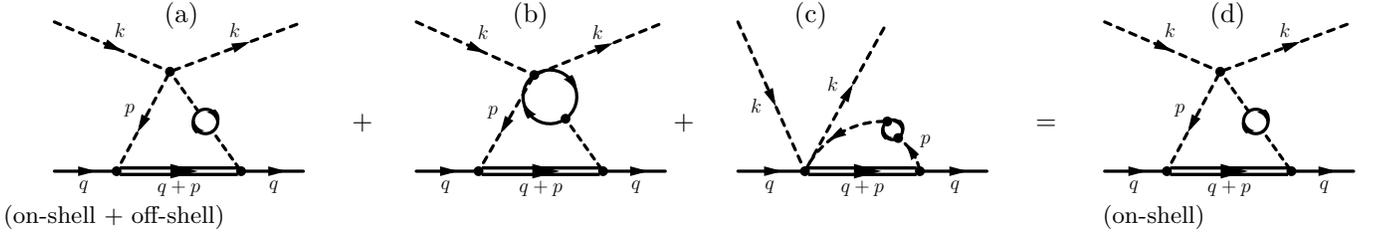}
\caption{Additional set of triangle diagrams with $\Delta(1232)$. The figure caption from Fig. \ref{fig:yet_another} also applies here.}
\label{fig:yet_another_delta_version}
\end{figure}
In Fig. \ref{fig:yet_another_delta_version} we plot the analogous diagrams of Fig. \ref{fig:yet_another} but with a $\Delta$ intermediate state, instead of a nucleon. 
The same type of off-shell cancellation found for the diagrams in Fig. \ref{fig:yet_another} holds also here as has been shown in \cite{Oset:2000gn,Kaskulov:2005kr} for a similar configuration.
This means, with $p$ $(k)$ being the loop momentum (external momentum): $p^2\to m_\pi^2$, $k^2\to m_\pi^2$, $pk\to p^0 k^0$ (the term $pk$ is not affected by the off-shell cancellation). In the last substitution, the integration over the spatial part ${\bf pk}$ vanishes.

Including a factor of two from inserting the medium correction in either pion line, the correction for $\pi^-p\to\pi^-p$ is
\be
\delta V^{(\ref{fig:yet_another_delta_version},\, d)}&=&-\frac{4\,f_{\pi N\Delta}^{*2}}{9\,f_\pi^2\,m_\pi^2}\int\limits_0^\infty\frac{dp\;p^2}{2\pi^2}\;F^2(p)\;\frac{p^2}{\eta}\non
&\times&\left(
-\frac{m_\pi^2(M_N-E_\Delta-2\eta)-2\eta^2\,k^0}{2\eta^2\,(M_N-E_\Delta-\eta)^2}+\int\limits_0^\infty d\omega\;\frac{S_\pi(\omega,p,\rho)}{\omega+\eta}\;\frac{2m_\pi^2\,(M_N-E_\Delta-\omega-\eta)-4\omega\eta\;k^0}{(M_N-E_\Delta-\eta)(M_N-E_\Delta-\omega)}\right).
\label{secfamdelta}
\ee
Here, $E_\Delta$ is the $\Delta(1232)$ energy. As an explicit calculation shows, the term with $2m_\pi^2$ inside the $\omega$-integral is of isoscalar nature whereas the term with $k^0$ is of isovector nature. This means that the term with $2m_\pi^2$ is the same for all channels of our coupled channel approach whereas one has to multiply $\delta V^{(\ref{fig:yet_another_delta_version},\, d)}$ by $-1$, $-\sqrt{2}$, and $0$ for $\pi^-n\to\pi^-n$, $\pi^-p\to\pi^0 n$, and $\pi^0n\to\pi^0n$, respectively. As nucleon and $\Delta$ mass are non-degenerate, the isoscalar part does not cancel as it had been the case for the diagrams of Fig. \ref{fig:yet_another}.

In Eq. (\ref{example_SRC}) we have already seen an example for additional SRC: Between the nucleon emitting a pion and the $ph$ or $\Delta h$ medium insertions in the pion propagator, there are also SRC. If this is the case for both ends of the pion propagator, the substitution of the pion propagator $D_\pi$ is given by the projection technique employed in Eq. (\ref{example_SRC}). In the triangle diagrams, there is only one side of the pion line affected and the additional SRC can be cast in a substitution of the in-medium pion propagator $D_{(1')}$ from Eq. (\ref{dprop}) according to
\be
D_{(1')}\to D_{(1')}\;\times\;\frac{1}{1-g'\left(\frac{D+F}{2f_\pi}\right)^2\,F^2(p)\;U}
\label{add_short_range_one}
\ee
as an explicit calculation shows. This does not affect the off-shell cancellation behavior discussed before. Remember that we always use the same Migdal parameter for $NN$ and $N\Delta$ SRC. Although the contributions from Eqs. (\ref{correctyet},\ref{secfamdelta}) are sizable, they are suppressed  by a factor of five by the additional SRC from Eq. (\ref{add_short_range_one}). In the final numerical results they introduce a small correction.

\subsection{Isovector correction from the NLO $\boldsymbol{\pi N}$ interaction}
\begin{figure}
\includegraphics[width=0.3\textwidth]{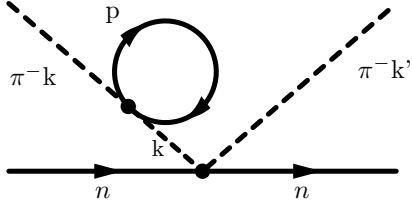}
\caption{Nucleon tadpole in the pion propagator with NLO chiral $\pi N$ interaction.}
\label{fig:another2}
\end{figure} 

The Weinberg-Tomozawa term is also renormalized by higher orders in the isoscalar $\pi N$ interaction. A correction of this type comes from the nucleon tadpole in the pion propagator shown in Fig. \ref{fig:another2}. The $\pi N$ interaction of the tadpole is from the $s$-wave isoscalar  interaction from the NLO chiral Lagrangian whereas the other $\pi N$ interaction is given by the Weinberg-Tomozawa term. The nucleon tadpole with isovector interaction vanishes in symmetric nuclear matter. 

The isoscalar $\pi N$ interaction in the NLO order is given by \cite{Meissner:1993ah}
\be
t_{\pi N}&=&\frac{4\,c_1}{f_\pi^2}\,m_\pi^2-\frac{2\,c_2}{f_\pi^2}\,(k^0)^2-\frac{2\,c_3}{f_\pi^2}\,k^2\non
&=&\left(\frac{4\,c_1}{f_\pi^2}\,m_\pi^2-\frac{2\,c_2}{f_\pi^2}\,\omega(k)^2-\frac{2\,c_3}{f_\pi^2}\,m_\pi^2\right)-\frac{2c_2+2c_3}{f_\pi^2}\,(k^2-m_\pi^2)=t_{\pi N}^{\rm on}+t_{\pi N}^{\rm off}.
\label{separated}
\ee
Here the interaction has been separated into on-shell and off-shell part \cite{Cabrera:2005wz} (the latter term with $(k^2-m_\pi^2)$). For the nucleon tadpole in Fig. \ref{fig:another2} and considering first the off-shell part, the selfenergy is given by $\Pi=-(2c_2+2c_3)(k^2-m_\pi^2)\rho/f_\pi^2$. The entire diagram is then given by
\be
V^{(\ref{fig:another2})}=-t_{\pi N\to\pi N}\,\frac{2c_2+2c_3}{f_\pi^2}\;(k^2-m_\pi^2)D(k)\rho
\ee
where $t_{\pi N\to\pi N}$ is the isovector interaction from the Weinberg-Tomozawa term and $D(k)$ the intermediate pion propagator which cancels the term $(k^2-m_\pi^2)$ from the isoscalar vertex. This means a vertex renormalization by a similar mechanism as we have already seen for diagram (4) in Fig. \ref{fig:more_higher} and
\be
\frac{\delta V^{(\ref{fig:another2})}}{V_{\rm WT}}=-\frac{2\,c_2+2\,c_3}{f_\pi^2}\,\rho.
\label{dvovervtad}
\ee
For the numerical evaluation we use the values of the $c$-coefficients of the fit $2^\dagger$ from Ref. \cite{Fettes:2000bb}, 
\be
c_1=-0.35\pm 0.1\,{\rm GeV}^{-1},\;c_2=-1.49\pm 0.67\,{\rm GeV}^{-1},\;c_3=0.93\pm 0.87\,{\rm GeV}^{-1}.
\label{cvalues}
\ee
It would be more consistent to use the values of the present fit in Tab. \ref{tab:parms} instead. However, in the present model we have only access to $c_2$ and the combination $2c_1-c_3$. In Eq. (\ref{dvovervtad}) the $c$-coefficients are combined in a different way, and we have to resort to the values of \cite{Fettes:2000bb}. In any case the values from Eq. (\ref{cvalues}) are compatible within errors with ours from Tab. \ref{tab:parms} (see also Ref. \cite{Doring:2004kt}). For an estimate of the theoretical error, we can also use the $c$-values from fit $2^*$ instead of $2^\dagger$ \cite{Fettes:2000bb}. This induces a theoretical error of the order of 20 \%
for the contribution which by itself is smaller than other diagrams. 

As for the on-shell part of the interaction in Eq. (\ref{separated}) we notice that the intermediate pion propagator, between the nucleon tadpole and the Weinberg-Tomozawa vertex, does not cancel. This means that the on-shell nucleon tadpole contributes to the pion selfenergy and not to the vertex renormalization and hence, if it is an external pion line, it will be automatically considered when solving the Klein Gordon equation for pions with the proper selfenergy and must not be included as a genuine new contribution. When this part of the tadpole occurs in internal lines, compared to the $p$-wave pion selfenergy, the $s$-wave selfenergy is small and can be neglected as we have also seen in the self consistent calculation in Sec. \ref{sec:selfcon}. 

\subsection{Results for the isovector renormalization}
\label{isovec_reno}
\begin{figure}
\includegraphics[width=0.4\textwidth]{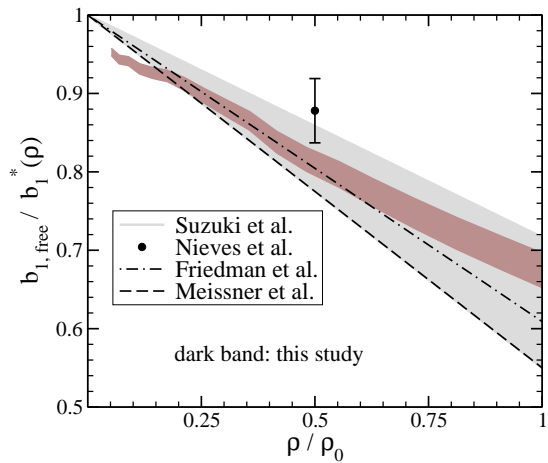}
\caption{In-medium isovector $b_1^*(\rho)$ compared to the vacuum isovector term $b_{1\,{\rm free}}$. The gray band from Suzuki {\it et al.} \cite{Suzuki:2002ae} is from a phenomenological fit, as well as the point from Nieves {\it et al.} \cite{Nieves:1993ev}. Also shown are chiral calculations from Mei\ss ner {\it et al.} \cite{Meissner:2001gz} and Weise {\it et al.}(Friedman {\it et al.}) \cite{Weise:2001sg,Friedman:2004jh}.}
\label{fig:c_44_renormalization}
\end{figure}

For all corrections evaluated in this section \ref{sec:higher_order}, the vertex corrections can be recast as a correction to the isovector interaction strength $b_1^*(\rho)$ or, in other words, an in-medium change of  $f_\pi$. Note that we refer to the $f_\pi$ that appears in the Weinberg-Tomozawa term of Eq. (\ref{kernel_vac}); we do not claim a universal change of $f_\pi$ in the nuclear medium (see also the caveat following Eq. (\ref{multif})).
For example, Eq. (\ref{multif}) gives the renormalization of $b_{1\,{\rm vac}}/b_1^*(\rho)$ from the diagrams (1) to (4) of Figs. \ref{fig:first_vertex},\ref{fig:more_higher} with an overall value of $r=1$. Including these diagram as well as all other isovector corrections found, the in-medium change of $b_1$ is plotted in Fig. \ref{fig:c_44_renormalization}. 

The result in Fig. \ref{fig:c_44_renormalization} is given at $\Lambda=0.9$ GeV for the monopole form factor that appears in the $ph$ and $\Delta h$ pion selfenergies of the vertex corrections (see Eqs. (\ref{migdal},\ref{selftranslong},\ref{add_short_range_one})). The dependence on $\Lambda$ is moderate and at $\Lambda=1$ GeV $b_1^*(\rho)/b_{1, {\rm free}}$ is increased by another 10 \%. With the decrease of $f_\pi$ in the medium the isovector $\pi N$ interaction effectively increases, in quantitative agreement with a recent analysis of deeply bound pionic atoms \cite{Suzuki:2002ae} and the phenomenological fit from \cite{Nieves:1993ev}
and the analysis of $\pi A$ scattering from \cite{Friedman:2004jh}. There is also qualitative agreement with other theoretical works \cite{Weise:2001sg,Meissner:2001gz} which are justified in a different way, without the thorough many body study done here.

In Fig. \ref{fig:c_44_renormalization} we give a band of values for our results of $b_1$, including the uncertainties discussed above from the diagram of Fig. \ref{firstdelta}, plus a 20 \% extra uncertainty from the dependence on the form factor. We can see that the band overlaps with the experimental band of \cite{Suzuki:2002ae}. 

After studying vertex corrections of isovector type in this section, next we turn to study vertex corrections of isoscalar type and their effects in the isoscalar optical potential.

\section{Renormalization of the NLO isoscalar term in $\pi N$ scattering}
\label{sec:renoiso}
The model from Ref. \cite{Doring:2004kt} for the $\pi N$ interaction in the vacuum has two sources for isoscalar contributions: one is the NLO, point-like, interaction from Eq. (\ref{iso_lagrangian}) and the other one comes from the isovector term in the rescattering of the pion generated in the Bethe-Salpeter equation. In fact, the latter is quite large, $b_g=442\cdot 10^{-4}\,m_\pi^{-1}$ (see Tab. VII from \cite{Doring:2004kt}). This large contribution is partly canceled by the NLO contact term from Eq. (\ref{iso_lagrangian}) that is $b_c=-336\cdot 10^{-4}\,m_\pi^{-1}$, leading to a final value of $b_0=-28\cdot 10^{-4}\,m_\pi^{-1}$.

For the application of the model in nuclear matter this partial cancellation has consequences. Renormalizing the isovector strength changes the in-medium isoscalar term through the rescattering piece. Then, the sum of this term and the point-like NLO interaction will not show the partial cancellation of the vacuum case any more. It is therefore important to treat the NLO isoscalar term on the same footing as the isovector renormalization. 

It is easy to see what the effects of the increase of $b_1$ will be in the isoscalar part of the potential. Indeed, if one cared only about Pauli blocking corrections in the intermediate nucleons, the effect would be given by Eq. (\ref{ericson_correction}) with an increased $b_1$, thus leading to an increased isoscalar repulsion. However, the $\pi N$ rescattering term with no Pauli blocking, which is larger, has opposite sign (see sign of $b_g$). Hence, the net effect of increasing $b_1$ in the medium, including Pauli blocking corrections, is a net attraction, with opposite results to a naive implementation of the $b_1$ changes in the Ericson-Ericson formula of Eq. (\ref{ericson_correction}). In some analyses of data \cite{Friedman:2004jh} the needed repulsion is obtained by using the Ericson-Ericson Pauli blocking correction, Eq. (\ref{ericson_correction}), with the increased $b_1$. This is one empirical way to implement repulsion, but from the theoretical point of view one should evaluate the full $\pi N$ $t$-matrix in the medium, not just the Pauli correction to it, and this means that the problem of the missing repulsion becomes more acute. 

The diagrams from the last section will be the guideline for the renormalization of the NLO isoscalar. We do not redraw these diagrams, but the $\pi N\to\pi N$ contact interaction is now given by the NLO isoscalar term instead of the LO Weinberg-Tomozawa isovector term.

\subsection{Tadpole and off-shell contributions}
\begin{figure}
\includegraphics[width=4cm]{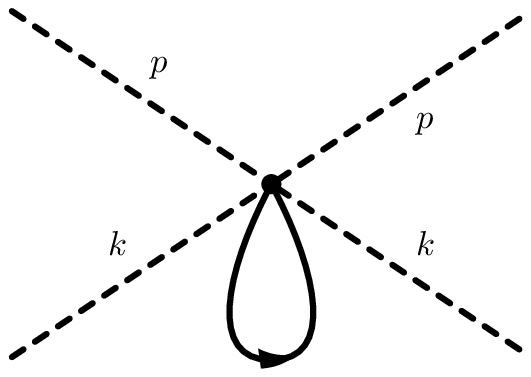}\hspace*{1cm}
\includegraphics[width=4cm]{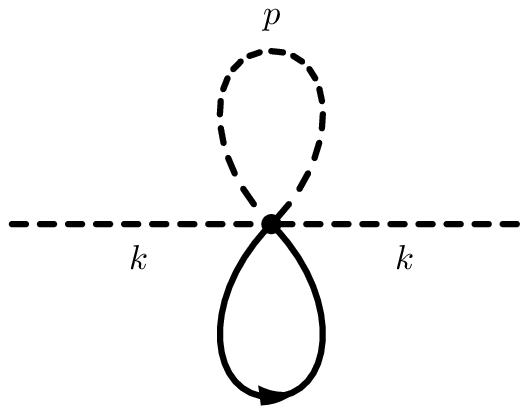}\hspace*{1cm}
\includegraphics[width=4cm]{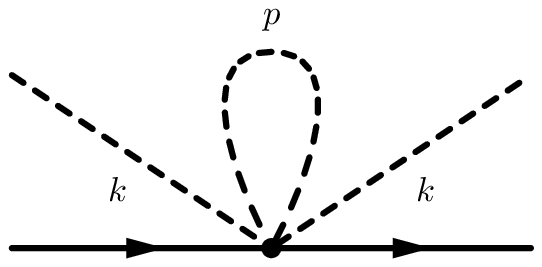}
\caption{Pion-pion interaction with a nucleon tadpole from the NLO $\pi N$ interaction (left). Closing one pion line produces a pion selfenergy of the $t\rho$ type (center). This $t$ is a $\pi N$ vertex correction (right), with the same geometry as diagram (1) in Fig. \ref{fig:first_vertex}.}
\label{fig:twotad}
\end{figure}

We start with the pion tadpole (1) from Fig. \ref{fig:first_vertex}. The NLO isoscalar Lagrangian has to be expanded to four pion lines in order to provide the $4\pi 2N$ vertex needed in this diagram. As shown in the following, to this end we can utilize the in-medium Lagrangian derived in Ref. \cite{Thorsson:1995rj,Wirzba:1995sh,Kirchbach:1996xy,Kirchbach:1997rk} (see also \cite{Cabrera:2005wz,Jido:2000bw}) by taking the mean-field approximation for the nucleon field. The terms with a medium correction $\rho$ of the nuclear density read 
\be
\langle {\cal L}\rangle=\frac{1}{2}\,\rho\left(c_3\,{\rm Tr}[\partial U\partial U^\dagger]+c_2\,{\rm Tr}[\partial_0 U\partial_0 U^\dagger]\,+c_1\,{\rm Tr}[U^\dagger\chi+\chi^\dagger U]\right)
\label{inmediso}
\ee
by keeping only the isoscalar terms which are parametrized in terms of $c_1,\,c_2,\,c_3$.
Expanding this term up to four external pion lines leads to a $\pi\pi$ vertex with a nucleon tadpole as displayed in Fig. \ref{fig:twotad} to the left. Contracting two of the pion fields leads to a diagram that appears as a pion selfenergy with a pion tadpole and a nucleon tadpole as displayed in the center of Fig. \ref{fig:twotad}. 
For a $\pi^-$ this selfenergy is given by 
\be
\Pi^{(\ref{fig:twotad})}=\frac{2\rho}{3f_\pi^4}\;i\int\frac{d^4p}{(2\pi)^4}\,D(p)\,\left(2c_3(k^2+p^2)+2c_2((k^0)^2+(p^0)^2)-\frac{5}{2}\,c_1\,m_\pi^2\right)
\label{firstt2tad}
\ee
with the pion propagator $D(p)$ and the momenta as assigned in Fig. \ref{fig:twotad}, center. The Lagrangian of Eq. (\ref{inmediso}), as well as its related one of Eq. (\ref{iso_lagrangian}), are meant to be used at very low pion energies and their extrapolation to high momenta is not justified. Actually, in the study of $\pi N$ of Ref. \cite{Doring:2004kt}, where an extrapolation to energies about 400 MeV above $\pi N$ threshold was done, the exponential damping factor of Eq. (\ref{iso_lagrangian}) was demanded by a fit to the data. Hence, we make here the sensible choice of setting $p^2=(p^0)^2=m_\pi^2$ in Eq. (\ref{firstt2tad}). While this certainly introduces some uncertainties, these are still small compared with larger sources of uncertainties that we shall discuss below.

Note the appearance of the nuclear density $\rho$ in Eq. (\ref{firstt2tad}): the selfenergy is of the type $t\rho$ with a matrix element $t$ that can be extracted by opening the nucleon line of the nucleon tadpole, meaning the division of Eq. (\ref{firstt2tad}) by $\rho$. This is displayed to the right in Fig. \ref{fig:twotad}. The resulting diagram is a $\pi N$ vertex correction with the same geometry as diagram (1) in Fig. \ref{fig:first_vertex} but using the NLO isoscalar interaction for the $4\pi 2N$ vertex. 

Next, we proceed like in the diagrams of Fig. \ref{fig:first_vertex} including $ph$, $\Delta h$, and short range correlations in the pion propagator of the third diagram of Fig. \ref{fig:twotad} and subtracting the free part of this $\pi N$ $t$-matrix. This gives a genuine many body correction, in the line of Eq. (\ref{multif}) but of isoscalar character.
As a result, the isoscalar vertex correction for the coupled channels $i,j$ reads 
\be
\delta V_{ij}^{(\ref{fig:twotad})}=\frac{\delta_{ij}}{3f_\pi^4}
\int \limits_0^\infty\frac{dp\;p^2}{2\pi^2}\;\int\limits_0^\infty d\omega\; \left(S_\pi(\omega,p,\rho)-\frac{\delta(\eta-\omega)}{2\eta}\right)\;\left[m_\pi^2\,(8c_3+4c_2-5c_1)+c_2(2k^0)^2\right].\non
\label{vijtwotad}
\ee
The in-medium correction from Eq. (\ref{vijtwotad}) has to be added to the kernel of the Bethe-Salpeter equation (\ref{bse}). The term with $S_\pi$ corresponds to the dressed pion propagator as in diagram (1') of Fig. \ref{fig:first_vertex} and the term with $\delta(\eta-\omega)$ to the vacuum diagram (1) which is subtracted. 

For the diagrams (2) and (3) from Fig. \ref{fig:more_higher}, one or two of the $\pi N$ vertices can be given by the NLO isoscalar interaction: the Bethe-Salpeter equation (\ref{bse}) iterates the kernel and allows for any combination of isoscalar (see Eq. (\ref{iso_lagrangian})) and isovector vertices (see Eq. (\ref{kernel_vac})) in the rescattering series. For two iterated Weinberg-Tomozawa vertices we have already determined the off-shell contributions from the direct and crossed term; for iterated isoscalar interactions we should do in principle the same. However, the contributions are smaller and we neglect them. This can be seen as following: the strength of the isoscalar interaction is $b_c=-336\cdot 10^{-4}\,m_\pi^{-1}$ \cite{Doring:2004kt} which is around one third of the isovector strength. A combination of two isoscalar vertices in $\pi N$ rescattering would, thus, approximately lead to an off-shell contribution nine times smaller than the off-shell effect from the combination of two isovector vertices studied before, and we can safely neglect it. A combination of one isoscalar vertex and one isovector vertex results in an overall isovector interaction and is of no interest in the present case where only symmetric nuclear matter is considered. 

The diagram (4) from Fig. \ref{fig:more_higher} renormalizes the NLO isoscalar interaction in the same way as it affected the isovector interaction studied before. This means $\delta V_{NLO}^{(4)}/V_{NLO}$ is given by the right hand side of Eq. (\ref{multif}), with $r=2/3$ as before, where $V_{NLO}$ is given by the second term in Eq. (\ref{iso_lagrangian}).

\subsection{Loop corrections in the $t$-channel}
In diagram (5) of Fig. \ref{fig:more_higher} the Weinberg-Tomozawa term can be replaced by the NLO isoscalar interaction. We will consider on-shell and off-shell part of this interaction, given in Eq. (\ref{separated}), separately. 

\subsubsection{On-shell part of the $\pi N\to\pi N$ vertex}
We consider the process $\pi^-p\to\pi^-p$ and a charged pion running in the loop. Then, the contribution from the on-shell part of the NLO Lagrangian reads (see first term of the right hand side of Eq. (\ref{separated}))
\be
-iT^{(5), {\rm on}}_{\rm NLO}&=&\int\frac{d^4p}{(2\pi)^4}\;(-it_{\pi N})\;iD(p)\;iD(p)\;(-it_{\pi\pi})\non
&=&\frac{1}{3f_\pi^2}\int\frac{d^4p}{(2\pi)^4}\;\frac{4c_1m_\pi^2-2c_3m_\pi^2-2c_2\eta^2}{f_\pi^2}\;\frac{1}{(p^2-m_\pi^2+i\epsilon)^2}\;(p^2+6pq+q^2-2m_\pi^2)
\label{dia5onshelliso}
\ee
with $\eta^2=m_\pi^2+{\bf p}^{\;2}$ and $D$ being the pion propagator as before. We have taken here already the limit of forward scattering. The last term comes from the $\pi\pi$ vertex with $q$ the momentum of the external pions. As we did before, we take the on-shell value for the external pions $q^2=m_\pi^2$ and can substitute the last term in Eq. (\ref{dia5onshelliso}) by $(p^2-m_\pi^2)$, taking into account that the mixed term $6pq$ vanishes due to symmetric integration. As in Sec. \ref{sec:dia5isovector} we dress only one of the propagators in order to stay in line with the other corrections evaluated (then, a multiplicity factor of 2 appears according to the two possibilities of inserting the in-medium correction in either pion propagator of the loop). The propagators in Eq. (\ref{dia5onshelliso}) are then given by $D^2\to D_{(1)}\,D_{(1')}$ in Eq. (\ref{dia5onshelliso}) for the medium part and $D^2\to D^2_{(1)}$ for the vacuum part with $D_{(1)},\,D_{(1')}$ from Eq. (\ref{dprop}). 

The NLO $\pi N$ amplitude in Eq. (\ref{dia5onshelliso}) has a term $2c_2\eta^2$ which introduces additional powers of $p$ in the integration. Although the integral is still convergent, the value of the $c_2$ coefficient is only valid for small momenta, where it has been determined in fits to low energy $\pi N$ scattering data (see Eq. (\ref{cvalues})). Once more, in analogy to what was done following Eq. (\ref{firstt2tad}), we replace $2c_2\eta^2\to 2c_2m_\pi^2$ in Eq. (\ref{dia5onshelliso}), taking thus the threshold value of the NLO isoscalar interaction. 

Furthermore, there is another off-shell cancellation, the one of the $\pi\pi$ vertex. This is due to the additional diagram shown in Fig. \ref{fig:twotad22}. In a similar way as in \cite{Oset:2000gn,Kaskulov:2005kr} (see also the diagrams in Fig. \ref{fig:yet_another}), the diagram cancels the off-shell part of the $\pi\pi$ interaction appearing in Eq. (\ref{dia5onshelliso}).

For the process $\pi^-p\to\pi^-p$, the pion in the loop can also be a $\pi^0$. 
It is easy to see that the overall correction, including all the intermediate pions with different charge, is of isoscalar type by explicitly calculating other $\pi N$ channels. 
Taking all this into account,
integrating the $p^0$-component, and subtracting the vacuum part from the medium part results in 
\be
\delta V^{(5), {\rm on}}_{\rm NLO}&=&-\frac{m_\pi^4}{f_\pi^4}\;\left(4c_1-2c_3-2c_2\right)\;\int\limits_0^\infty\frac{dp\;p^2}{2\pi^2}\;\frac{1}{\eta}\left[-\frac{1}{4\eta^2}+\int\limits_0^\infty d\omega\;\frac{S_\pi (\omega,p,\rho)}{\eta+\omega}\right].
\ee
This is a tiny correction to the isoscalar renormalization which is neglected in the final numerical results.

\subsubsection{Off-shell part of the $\pi N\to\pi N$ vertex}
The off-shell part of the NLO isoscalar interaction is renormalized in a similar way as before. Taking the term with $k^2-m_\pi^2$ of Eq. (\ref{separated}), the vacuum amplitude for $\pi^-p\to\pi^-p$ is in this case given by
\be
-iT^{(5), {\rm off}}_{\rm NLO}=-2\;\frac{2c_2+2c_3}{f_\pi^2}\frac{1}{3f_\pi^2}\int\frac{d^4p}{(2\pi)^4}\;\frac{1}{(p^2-m_\pi^2+i\epsilon)^2}\;(p^2-m_\pi^2)\left[(p^2+6pq+q^2-2m_\pi^2)+(p^2+q^2-\frac{1}{2}\;m_\pi^2)\right]
\ee
where the external pions are again at momentum $q$ and $(p^2-m_\pi^2)$ is the off-shell part of the NLO isoscalar vertex which cancels one of the propagators. 
In the square brackets, the contributions from having a charged pion or a neutral one are denoted separately.
It is easy to see that the overall contribution is again of isoscalar nature.

In the term $(p^2+6pq+q^2-2m_\pi^2)$ from the $\pi\pi$ vertex, the mixed product $pq$ vanishes due to symmetric integration. 
The $p^0$-integration is straightforward for both vacuum and medium loop. The resulting medium correction reads
\be
\delta V^{(5), {\rm off}}_{\rm NLO}&=&-4\;\frac{2c_2+2c_3}{3f_\pi^4}\int\limits_0^\infty\frac{dp\;p^2}{2\pi^2}\left[-\frac{3m_\pi^2}{8\eta}+\int\limits_0^\infty d\omega\;S_\pi(\omega,p,\rho)\left(\omega^2-p^2-\frac{1}{4}\;m_\pi^2\right)\right].
\label{offshellnlo}
\ee
The factor $\left(\omega^2-p^2-m_\pi^2\right/4)$ comes from on- and off-shell part of the $\pi\pi$-vertex. As before, the off-shell part cancels with the diagram from Fig. \ref{fig:twotad22}. Keeping only the on-shell part of the $\pi\pi$ interaction, the final result can be easily obtained and is given in Eq. (\ref{offshellnlo}) with the replacement
\be
\left(\omega^2-p^2-\frac{1}{4}\;m_\pi^2\right)\to \frac{3}{4}\;m_\pi^2.
\label{simplific1}
\ee

\subsection{Further renormalizations of the isoscalar $\pi N$ interaction}
Next, the diagram from Fig. \ref{fig:another} is considered which is now given by two $\pi NN$ vertices and one $\pi^-N\to\pi^-N$ transition from the NLO isoscalar interaction. The only change with respect to the previous result from Eq. (\ref{example_SRC}) is a change in isospin factors. Indeed, consider the $\pi^-p\to\pi^-p$ amplitude and the loops with $\pi^0p$ or $\pi^+n$ (see Fig. \ref{fig:another}), implying the $\pi N\to\pi N$ vertex to be $\pi^-p\to\pi^-p$ or $\pi^-n\to\pi^-n$, respectively. The isovector $\pi N$ vertex has opposite sign in these two cases while it is the same one for the isoscalar amplitude. Hence, the contributions of the two loops add in the case of the isoscalar correction while they get subtracted  in the case of the isovector correction. Thus we obtain $\delta V_{\rm NLO}^{(\ref{fig:another})}/V_{\rm NLO}=-3\,\delta V^{(\ref{fig:another})}_{\rm SRC}/V_{\rm WT}$ with $\delta V^{(\ref{fig:another})}_{\rm SRC}/V_{\rm WT}$ given by Eq. (\ref{example_SRC}). 

Considering the nucleon tadpole in Fig. \ref{fig:another2} it is clear that this correction renormalizes in the same way the isoscalar and the isovector interaction because the tadpole  factorizes with the $\pi N$ amplitude. Thus, the renormalization of the isoscalar amplitude $\delta V_{\rm NLO}^{(\ref{fig:another2})}/V_{\rm NLO}$ is again given by the right hand side of Eq. (\ref{dvovervtad}).

When renormalizing the Weinberg-Tomozawa interaction, we have also considered the diagrams in Fig. \ref{fig:bull}. They have been found small as argued at the end of Sec. \ref{sec:renotomo} due to the occurrence of the difference of Lindhard functions $\bar{U}(q^0+k^0)-\bar{U}(q^0-k^0)$. The minus sign was a consequence of the isovector nature of the Weinberg-Tomozawa vertex. However, if one replaces the Weinberg-Tomozawa vertices at the bottom of the diagrams in Fig. \ref{fig:bull} with isoscalar ones, one obtains the combination $\bar{U}(q^0+k^0)+\bar{U}(q^0-k^0)$. This might result in a significant contribution. However, as the following argument shows, we should not consider this contribution as it would mean double counting: We consider the diagrams in Fig. \ref{fig:bull} with the nucleon line closed. This corresponds to a contribution to the $s$-wave pion selfenergy. However, the resulting selfenergy, let it be $\Pi_{(\ref{fig:bull})}$, is already generated in a different piece: Imagine the one-loop rescattering of two isovector interactions. Insert now a nucleon tadpole in the intermediate pion line as displayed in Fig. \ref{fig:another2}. Joining the external nucleon lines in this rescattering diagram generates again the pion selfenergy $\Pi_{(\ref{fig:bull})}$. For similar reasons of double counting we also discard vertex corrections that occur when one replaces the vertices at the bottom of Fig. \ref{fig:bull} with the NLO isoscalar interaction and, additionally, also the other $\pi N$ vertices.

Finally, let us remember there is a small vertex correction to the isoscalar interaction from the triangle diagrams with intermediate $\Delta$ of Fig. \ref{fig:yet_another_delta_version} which we have already discussed and evaluated in Sec. \ref{sec:triangle_delta}, Eqs. (\ref{secfamdelta},\ref{add_short_range_one}). 

\begin{figure}
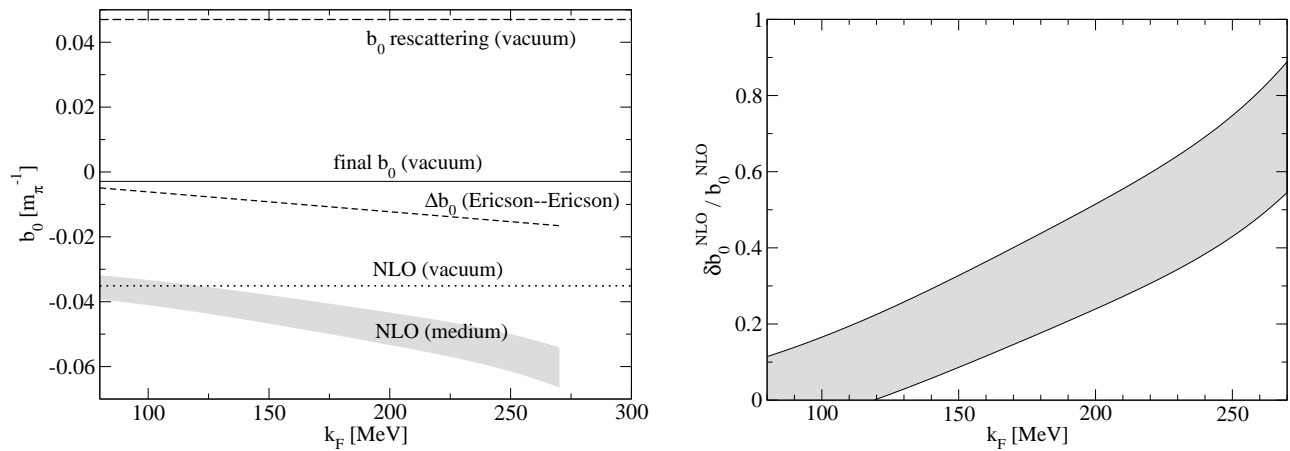

\includegraphics[width=8.5cm]{fig191_op.eps}\hspace*{0.5cm}
\includegraphics[width=7.9cm]{fig192_op.eps}
\caption{In-medium isoscalar interaction. Left: different sources of isoscalar contributions are shown (see text). Right: The fraction $\delta b_0^{\rm NLO}/b_0^{\rm NLO}$, actually calculated here, is plotted as a function of $k_F$.}
\label{fig:renoiso}
\end{figure}

The sum of the isoscalar corrections calculated in this section results in an increase of the NLO interaction as shown in Fig. \ref{fig:renoiso}.
On the right hand side, the $\delta b_0^{\rm NLO}$, obtained from the various corrections $\delta V_{NLO}$ of this section, is shown, divided by $b_0$ from the vacuum NLO isoscalar term given by the second term of Eq. (\ref{iso_lagrangian}). The ratio is of similar size as the isovector case shown in Fig. \ref{fig:c_44_renormalization}. On the left hand side of Fig. \ref{fig:renoiso} we show again the NLO isoscalar $b_0$ term for vacuum and medium. Additionally, the vacuum isoscalar term from the rescattering of isovector interactions is plotted, indicated with ''$b_0$ rescattering (vacuum)''. This illustrates how these two sources of isoscalar contribution almost cancel in the vacuum. When the full $\pi N$ vertices are used in the rescattering, the small ''final $b_0$ (vacuum)'' appears, as required by the vacuum data \cite{Doring:2004kt}. Fig. \ref{fig:renoiso} also shows, that the isoscalar in-medium vertex corrections, evaluated in this section, provide a large source of repulsion (negative $b_0$), even larger than the Ericson-Ericson correction from Eq. (\ref{ericson_correction}), which is also shown in the figure. In Sec. \ref{theo_unc} we will test the final results of the medium calculation for changes of the vacuum model. In particular, we will perform refits requiring a smaller NLO isoscalar vacuum term, which then also reduces the size of isoscalar contribution  from rescattering (see Fig. \ref{fig:renoiso}). However, as will be seen in Sec. \ref{theo_unc}, the final results are stable under these variations.

\section{Methodology of the expansion}
\label{sec:ordering}
The diagrams introduced follow the standard approach of field theory in which selfenergy corrections as well as vertex corrections are introduced in a perturbative expansion. Here, we want to discuss the methodology behind the expansion done. Let us proceed step by step. The first step consisted in developing an accurate model for the free $\pi N$ scattering amplitude in Sec. \ref{sec:vacuum}. There, we have followed the standard procedure of the chiral unitary approach. The Bethe-Salpeter (BS) equation is used to ensure unitarity with a kernel derived from chiral Lagrangians in which the lowest order and next-to-lowest Lagrangians are considered. An expansion in powers of momenta of the pion is done in the kernel of the BS equation, and the BS equation resums the higher order terms, like in the Schr\"odinger equation starting with a potential, ensuring the unitarity of the amplitude. 

When this is done, the many body expansion comes into action. Attempts of a systematic many body expansion along standard lines of chiral perturbation theory have been made in \cite{Meissner:2001gz}, but one automatically faces problems ultimately tied to the fact that the many body corrections come from $ph$ excitations which have small energy (starting from zero in our Fermi sea). This is in contrast with chiral perturbation theory where the success, for instance in $\pi\pi$ scattering, is tied to the large energy gap between the $\pi$ and the next meson mass excitation, the one of the $\rho$. Even in $\pi$-deuteron scattering many options of expansion parameters are suggested \cite{Beane:2002wk} with quite different results among them. 

\begin{figure}
\includegraphics[width=0.3\textwidth]{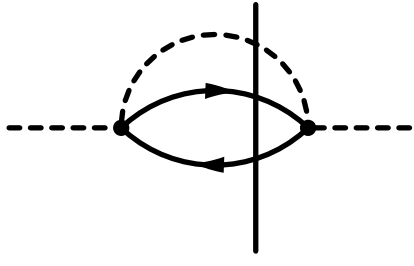}
\caption{Pion self energy for one $\pi N$ loop.}
\label{fig:cut_dia}
\end{figure}

\begin{figure}
\includegraphics[width=0.3\textwidth]{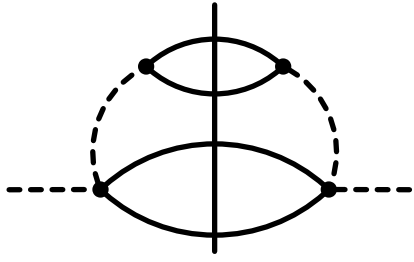}
\caption{Two-nucleon pion absorption.}
\label{fig:twon}
\end{figure}

Our guiding line for the many body expansion follows the traditional many body approach in nuclear physics where one relies upon an expansion in powers of the density. The density is assumed to be a small quantity and this is based in phenomenological facts. Let us illustrate this in our case of $\pi$-nucleus scattering. Let us take Fig. \ref{fig:inmedium_closed} where the selfenergy of the pion is obtained by summing the $\pi N$ $T$-matrix over the occupied states of the Fermi sea. Technically, we have introduced a hole line in the diagram which induces a power of $\rho$ upon integration over the occupied states. Let us now cut the diagram by a vertical line, cutting the pion and the hole line. Restricting ourselves to one meson-baryon loop for simplicity, we have the result of Fig. \ref{fig:cut_dia}. 
The whole sum of loops would replace the $\pi\pi NN$ vertex by the full $\pi N$ $T$-matrix. Placing on-shell the intermediate states,  cut by the vertical line in Fig. \ref{fig:cut_dia}, as it would be implemented using Cutkowsky rules \cite{Carrasco:1989vq}, one accounts for $\pi$ transitions to a $\pi+(ph)$ channel. In other words, the diagram accounts for $\pi N\to \pi N$, which is the quasielastic scattering. Let us now dress the intermediate pion by letting it excite a $ph$, like in Fig. \ref{fig:twon},
and let us now cut the two $ph$ excitations by a vertical line placing them on-shell. The process that this cut accounts for is what is called genuine two-body absorption. One has $\pi NN\to NN$ and the pion disappears. Experimentally one has, for energies not close to threshold, bigger quasielastic cross sections than absorption ones. Close to threshold, Pauli blocking makes the quasielastic scattering small, which means that the imaginary part of the diagram of Fig. \ref{fig:cut_dia} becomes small, but this is not the case for the real part which is not subjected to the Pauli exclusion principle \cite{FeWa1971} and which has the size of the imaginary part when Pauli blocking does not restrict it. Much work has been done in the past on three body absorption, both experimentally \cite{Weyer:1990ye} and theoretically \cite{Oset:1986yi}, obtaining smaller rates than for two-body absorption, particularly at energies close to threshold. Thus, an expansion in the number of $ph$ lines, or in general terms, a hole line expansion, is the guiding principle for our many body approach, the justification for it coming from phenomenology. 

\begin{figure}
\includegraphics[width=1\textwidth]{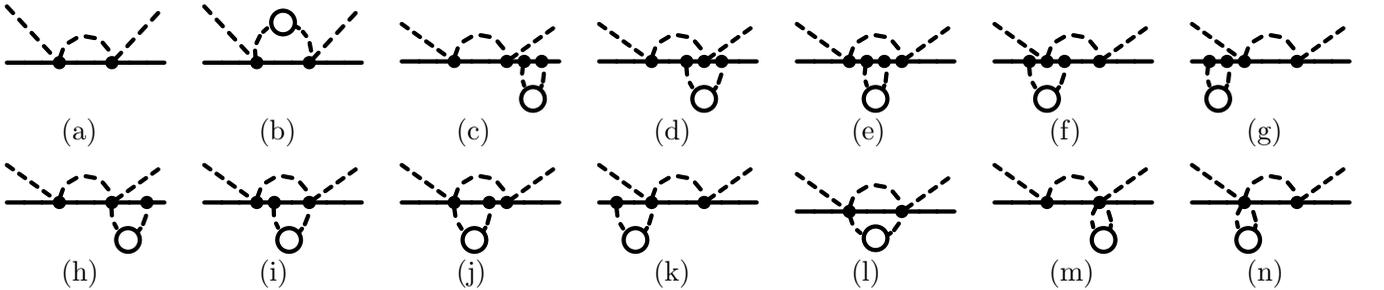}
\caption{Systematical $ph$ insertions in the basic rescattering loop.}
\label{fig:system}
\end{figure}

\begin{figure}
\includegraphics[width=0.3\textwidth]{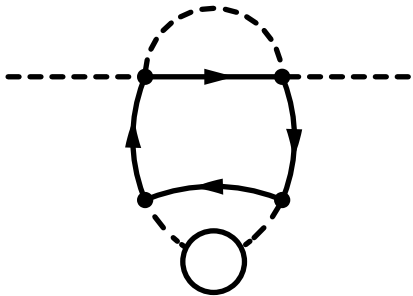}
\caption{Hole self energy.}
\label{fig:folded_hole}
\end{figure}

This said, let us see what a strict bookkeeping along the lines described above would give us when we introduce an extra $ph$ in the basic diagrams. Start from the  loop diagram of Fig. \ref{fig:system}(a)
 and introduce a new $ph$. Fig. \ref{fig:system}(b) is accounted for by our diagram of Fig. \ref{fig:inmedium}. Fig. \ref{fig:inmedium}(c) plus \ref{fig:inmedium}(g), external line renormalizations, introduce a selfenergy in the nucleon line when folded and converted into a hole line like in Fig. \ref{fig:folded_hole}.
This has to be considered together with diagram (e) in the sense that both particle and hole have to be renormalized simultaneously to account for the selfenergy in the medium. Yet, this selfenergy, ignoring non-localities, as done in our approach, gets cancelled in the $ph$ propagator $[E_\pi+E_h-E_p]^{-1}$. The diagrams (d) and (f) have already been accounted for in the diagram of Fig. \ref{fig:another}. Diagrams (h), (i), (j), (k) have already been considered before in Fig. \ref{fig:yet_another}. Similarly, diagrams (m) and (n) are taken into account by the vertex correction from Fig. \ref{fig:first_vertex}. We are left as novel diagram with the one of Fig. \ref{fig:inmedium}(l) which involves the $\pi\pi N$ intermediate states that were considered in \cite{Inoue:2001ip} and found to be relevant for $I=3/2$ at  energies around $s^{1/2}=1.5$ GeV, very small for $I=1/2$ at these energies, and negligible at $\pi N$ threshold that we study here, as we noted at the end of Sec. \ref{sec:full_model}. We use this fact here to neglect the medium modifications to this negligible term in free space. 

The discussion presented before clarifies further the procedure followed in former sections, where following a traditional approach in field theory one studies selfenergy and vertex corrections. 

In addition to this, our approach introduced further $ph$ excitations following again traditional lines in field theory and many body. For the selfenergy corrections, see Fig. \ref{fig:inmedium}, the $ph$ excitation is induced as a selfenergy of the pion, and thus, iterated $ph$ excitations in the Dyson sense are automatically accounted for. This is done technically with no extra effort and produces the coupled branches ($\pi$ and $ph$, or $\Delta h$) of the pion in the medium \cite{Aouissat:1994sx}.

The self consistency that we have implemented, introducing the full calculated $\pi$ selfenergy into the intermediate $\pi$ of Fig. \ref{fig:inmedium} also accounts for the $s$-wave self energy of higher order and is always desirable in many body calculations, although its effects here are moderate. This is unlike the case found in the $K^-$-nucleus interaction \cite{Ramos:1999ku} where the presence of a resonance close below threshold renders this procedure rather important. 

The discussion in this section has clarified further the principles followed in the many body expansion, which are ultimately based on phenomenological facts and not in a formal expansion based on one single parameter, as one has in some theories for elementary particle interactions.

\section{Numerical results}
\label{sec:numres2}
In the last sections \ref{sec:higher_order} and \ref{sec:renoiso}, vertex corrections for both the isovector and the isoscalar interaction have been evaluated. 
Together with the in-medium $\pi N$ loops $G_{\pi N}$, shown in Eq. (\ref{matterg}), we obtain a new $\pi N\to\pi N$ transition $T$ in Eq. (\ref{bse}). Integrating over the nucleons of the Fermi seas according to Eq. (\ref{sum_fermi}), the $s$-wave pion selfenergy is evaluated. The results are for symmetric nuclear matter as the calculations from Secs. \ref{sec:higher_order} and \ref{sec:renoiso} are performed in this limit. 

The contributions to the $s$-wave pion self energy can be ordered in powers of the density. In the following, we discuss the contributions up to order $\rho^2\rho^{1/3}$ for the external pion $s$-wave selfenergy $\Pi_S$. In Sec. \ref{theo_unc}, we compare this to the result up to all orders in $\rho$. In the $\pi N$ rescattering loops, the pion $p$-wave polarization accounts for one power of $\rho$ whereas the closing of the external nucleon line, as indicated in Fig. \ref{fig:inmedium_closed}, corresponds to another power of $\rho$. The Pauli blocking of the intermediate nucleon in the $\pi N$ rescattering generates an extra power of $\rho^{1/3}$ as can be seen in Eq. (\ref{ericson_correction}). Of course, higher powers are also contained: from the multiple rescattering generated by the BSE equation (\ref{bse}) on one hand, and the resummation of $ph$, $\Delta h$ pion selfenergies from Eq. (\ref{migdal}) on the other hand; however, these higher order corrections are small. 

Another contribution to $\Pi_S$ comes from the vertex diagrams from Secs. \ref{sec:higher_order}, \ref{sec:renoiso}, which are of order $\rho$ through the $p$-wave pion polarization. Closing the nucleon line (order $\rho$) for these diagrams, corrections of order $\rho^2$ are obtained. The vertex diagrams induce also corrections at $\rho^2\rho^{1/3}$: those consist of the Ericson-Ericson rescattering piece from Fig. \ref{fig:inmedium} {\it without} pion polarization of the rescattered pion, but with vertex corrections from Secs. \ref{sec:higher_order} and \ref{sec:renoiso} for exactly one of the Weinberg-Tomozawa $\pi N$ interactions. Some examples can be seen in Fig. \ref{fig:system}. These contributions are most easily included by multiplying the rescattering term by $(b_1^*(\rho)/b_{1,\,{\rm free}}-1)$, with $b_1^*(\rho)/b_{1,\,{\rm free}}$ from Fig. \ref{fig:c_44_renormalization}.

\begin{figure}
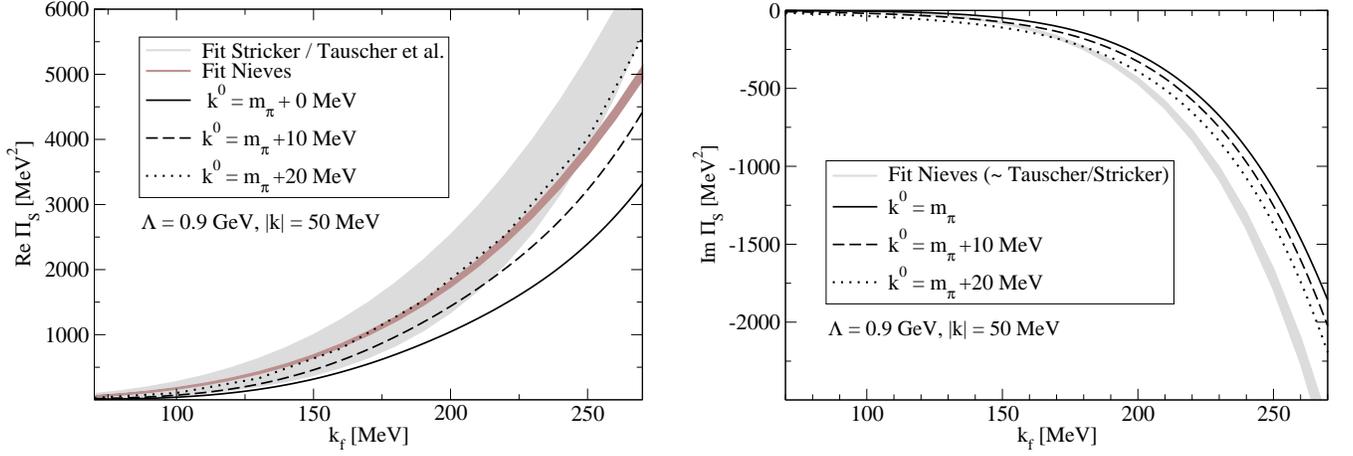

\includegraphics[width=8.5cm]{fig241_op.eps}\hspace*{0.5cm}
\includegraphics[width=8.5cm]{fig242_op.eps}
\caption{The $s$-wave pion selfenergy in nuclear matter up to order $\rho^2\rho^{1/3}$, for three pion energies $(k^0=m_\pi, \,k^0=m_\pi+10\,{\rm MeV},\,k^0=m_\pi+20\,{\rm MeV})$. The gray band shows the area of the phenomenological fits from Refs. \cite{Stricker:1980vm,Tauscher:1974bx} and the dark band the fit from \cite{Nieves:1993ev}. For ${\rm Im}\Pi_S$ all phenomenological fits lie in the gray shaded area.}
\label{fig:finalnum1}
\end{figure}

Summing all contributions, the pion $s$-wave self energy up to order $\rho^2\rho^{1/3}$ is plotted in Fig. \ref{fig:finalnum1} with the black solid lines. For the pion three-momentum, we have taken a typical value of $|{\bf p}|=50$ MeV although $\Pi_S$ depends only weakly on ${\bf p}$. The gray band shows the area of the experimental fits to pionic atoms from Refs. \cite{Stricker:1980vm,Tauscher:1974bx} (see also Fig. \ref{fig:Pi_overview}) whereas the dark band represents the phenomenological fit from \cite{Nieves:1993ev}. The present result for the external pion energy $k^0=m_\pi$ stays some 30 \% below the phenomenological fit. 
Note that at the order $\rho^2\rho^{1/3}$ considered here, the imaginary part ${\rm Im}\Pi_S$ is the same as in Fig. \ref{fig:im_Pi_overview}.

In Refs. \cite{Ericson:1981hs}, \cite{Kolomeitsev:2002gc}, \cite{Friedman:2007qx} it has been claimed, that a possible way of understanding the repulsion in pionic atoms comes from the energy dependence of the pion self energy; a consistent treatment of the Coulomb potential in the Klein-Gordon equation requires that the argument of the pion self energy is $\Pi_S(\omega-V_c)$ \cite{Kolomeitsev:2002gc} rather than $\Pi_S(\omega=m_\pi)$. Furthermore, the small isoscalar $\pi N$ potential at threshold rises rapidly with increasing energy and, thus, large effects from the energy dependence of $\Pi_S$ can be expected. 

In order to see this effect in the present calculation we have plotted $\Pi_S$ also for $p^0=m_\pi+10$ MeV and $p^0=m_\pi+20$ MeV (dashed and dotted line, respectively). The Coulomb potential for a nucleus with $A=100, \,Z=50$ can reach $V_c\sim 16$ MeV at an effective density of $\rho=\rho_0/2$, and even more for heavier nuclei. As Fig. \ref{fig:finalnum1} shows, the energy dependence leads indeed to an extra repulsion which agrees well with the phenomenological fits. However, theoretical uncertainties are larger than thought as will be discussed in Secs. \ref{theo_unc}, \ref{sec:roper}.

\subsection{Theoretical uncertainties}
\label{theo_unc}
\begin{figure}
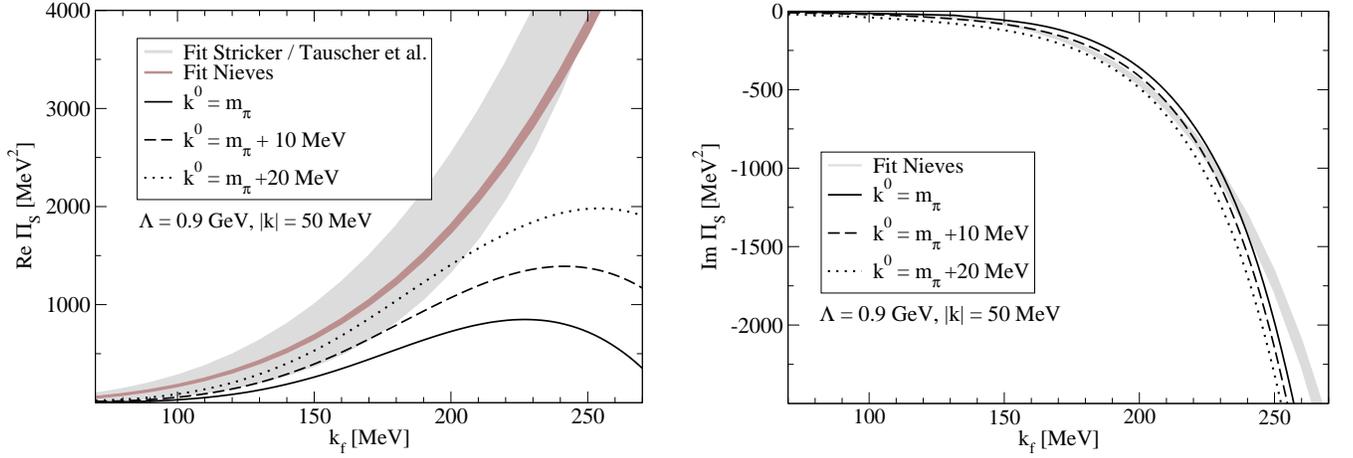

\includegraphics[width=8.5cm]{fig251_op.eps}\hspace*{0.5cm}
\includegraphics[width=8.5cm]{fig252_op.eps}
\caption{$s$-wave pion selfenergy to all orders in $\rho$. Different pion energies $(k^0=m_\pi, \,k^0=m_\pi+10\,{\rm MeV},\,k^0=m_\pi+20\,{\rm MeV})$. Phenomenological fits as in Fig. \ref{fig:finalnum1}. }
\label{fig:finalnum2}
\end{figure}

Certain higher order corrections in density play an important role but at densities higher than felt by pionic atoms. E.g., at order $\rho^3\rho^{1/3}$, diagrams appear where both isovector vertices of the large rescattering piece contain the corrections from Sec. \ref{sec:higher_order}. These corrections reduce Re $\Pi_S$ from Fig. \ref{fig:finalnum1}. Dressing all vertices with the corrections found and using the $\pi N$ loop function with Pauli blocking and pion polarization, i.e. including all corrections found, to all orders, gives the result indicated in Fig. \ref{fig:finalnum2}.
Changes with respect to Fig. \ref{fig:finalnum1} are mainly due to the fact, that the vertex corrections can occur quadratically and higher, while before, at order $\rho^2\rho^{1/3}$, only one vertex correction enters the rescattering series.

The imaginary part on the right hand side in Fig. \ref{fig:finalnum2} is more negative and closer to the values from phenomenological fits: the imaginary part comes from the $ph$ insertions in the $\pi N$ loop function. The imaginary part of this rescattering loop is enhanced by the larger strength in both $\pi N$ vertices due to the vertex corrections.

The decrease of the real part can be understood as following: in the vacuum model the rescattering piece introduces an attraction which is compensated by a repulsion $b_c$ from the NLO isoscalar interaction at tree level \cite{Doring:2004kt}. The isovector interaction in the medium is increased from vertex corrections as we have seen in Fig. \ref{fig:c_44_renormalization}; this leads to an increase of the attraction from the rescattering piece. The Pauli blocking of the intermediate nucleon, namely the Ericson-Ericson effect from Eq. (\ref{ericson_correction}), which is repulsive, can not fully compensate this effect; as a result, the net repulsion is smaller than without vertex corrections. 
However, for arguments of external pion energies of $k^0=m_\pi+10$ MeV, $k^0=m_\pi+20$ MeV due to Coulomb shift, the results are in the region of the required repulsion at $k_F\sim 210$ MeV ($\rho=\rho_0/2$). It should be noted that the net attraction of the the increased $b_1$ in the medium, through rescattering, is a rather large effect, which more than compensates for the enhanced isoscalar amplitude in the medium found in Sec. \ref{sec:renoiso} and shown in Fig. \ref{fig:renoiso}.

We have tested the stability of our results. The corrections discussed stemming from rescattering are tied to our elementary vacuum model of Ref. \cite{Doring:2004kt}. For instance, the size of the isoscalar contribution $b_c$ from the NLO Lagrangian is correlated with the subtraction constant of the $\pi N$ vacuum loop, $a_{\pi N}$, in the fit. We have performed a refit of the vacuum amplitude requiring a smaller value of $b_c$. Close to threshold, a sufficiently good fit can be obtained with $b_c=-46 \cdot 10^{-4}\,m_\pi^{-1}$ which is around ten times smaller than $b_c$ from the fit of Tab. \ref{tab:parms}. It is interesting to note that the resulting parameter values $2c_1-c_3=-1.43$ GeV$^{-1}$, $c_2=-1.54$ GeV$^{-1}$ are quite close to the values from \cite{Fettes:2000bb} of $2c_1-c_3=-1.63\pm 0.9\,{\rm GeV}^{-1}$ and $c_2=-1.49\pm 0.67\,{\rm GeV}^{-1}$. However, even with this drastic change of the vacuum model, Re $\Pi_S$ hardly changes and the results are stable in this respect.

Further theoretical uncertainties come from the regularization scale $\Lambda$ that appears in the monopole form factors of the pion $p$-wave polarization. The result depends on $\Lambda$; a smaller value than the one used of $\Lambda=0.9$ GeV would provide slightly larger repulsion as we noted before. Nevertheless, the good agreement with the phenomenological analysis on the $b_1$ renormalization from \cite{Suzuki:2002ae}, which has been noted in Sec. \ref{isovec_reno}, provides support for this value of $\Lambda$.

We have also treated the pion self energy selfconsistent as in Sec. \ref{sec:selfcon}, including the $k^0$ energy dependence of the $s$-wave potential. This leads only to a 10 \% increase of Re $\Pi_S$ at $\rho=\rho_0/2$.

\subsection{Uncertainties from the Roper resonance}
\label{sec:roper}
There is another type of medium effect which has not been considered so far and will introduce additional uncertainties.
This is related to the Roper excitation and its decay into nucleon and two pions in $I=0$ and $s$-wave. The Roper is the lightest resonance with the same quantum numbers as the nucleon and allows for a decay into a nucleon and two pions which are in isospin zero and $s$-wave relative to each other and also relative to the nucleon. In Ref. \cite{Alvarez-Ruso:1997mx} the mechanism of Roper excitation from an isoscalar source and subsequent decay into two pions has been found dominant at low energies in the $NN\to NN\pi\pi$ production for pions in $I=0$. The isoscalar source can be described by an effective $\sigma$ exchange $\sigma NN^*$ between the nucleons, whose strength has been fitted independently for the $(\alpha,\alpha')$ reaction on a proton target \cite{Hirenzaki:1995js}. Based on that finding, the relevance of this mechanism in $\pi d$ scattering at low energies was also stressed in \cite{Meissner:2005ne}. We shall also consider it here in connection with the $s$-wave pion-nucleus optical potential.

\begin{figure}
\begin{flushleft}
\includegraphics[width=1\textwidth]{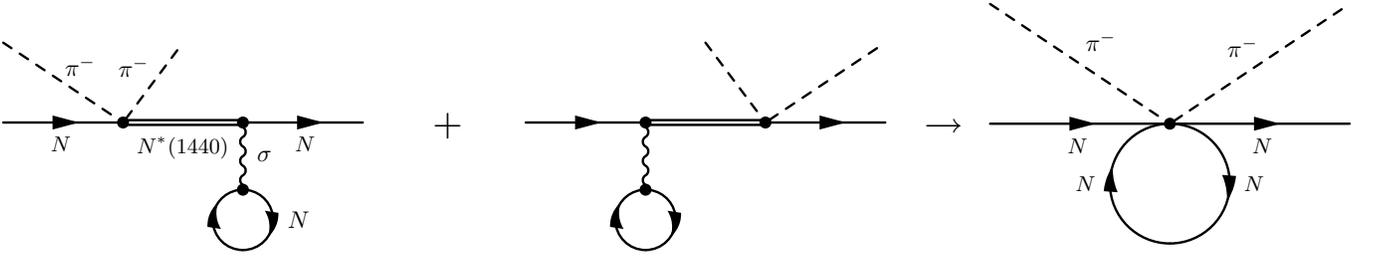}
\end{flushleft}
\caption{The Roper resonance in isoscalar, $s$-wave $\pi N$ scattering in the medium. On the right hand side the interaction in the heavy baryon limit is shown.}
\label{fig:roper}
\end{figure}

For the present purposes the mechanism described above can be adapted by having the two pions, one in the initial state and the other one in the final state, as indicated in Fig. \ref{fig:roper} on the left hand side. As the two pions are in a relative $I=0$ state, we obtain an isoscalar contribution to the $\pi N$ amplitude. The second nucleon line to which the isoscalar $\sigma$ couples is closed and gives a medium contribution to $\pi N$ scattering. In the heavy baryon limit the diagram reduces to a point-like interaction of a pion with two nucleons as indicated in Fig. \ref{fig:roper} on the right hand side.

For the $N^*N\pi\pi$ coupling an effective Lagrangian from Ref. \cite{Bernard:1995gx} is used which leads to the effective vertex \cite{Alvarez-Ruso:1997mx}
\be
-i\delta\tilde{H}_{N^*N\pi\pi}=-2i\,\frac{m_\pi^2}{f_\pi^2}\left(c_1^*-c_2^*\,\frac{\omega_1\omega_2}{m_\pi^2}\right)
\ee
for $\pi^+\pi^+$, $\pi^-\pi^-$ and $\pi^0\pi^0$ and zero otherwise (note a minus sign in $c_2^*$ with respect to \cite{Alvarez-Ruso:1997mx} because now one of the pions is incoming). Here $\omega_1,\omega_2$ are the energies of the pions. The values for the couplings are obtained in Ref. \cite{Alvarez-Ruso:1997mx} from a fit to the experimental width of the $N^*$ decay into $N\pi^+\pi^-$ and $N\pi^0\pi^0$, $c_1^*=-7.27\,{\rm GeV}^{-1}$ and $c_2^*=0\,{\rm GeV}^{-1}$. For the $N^*\sigma N$ coupling the effective vertex is $-i\Delta \tilde{H}_{\sigma NN^*}=iF(q)g_{\sigma NN^*}$ where $g_{\sigma NN^*}^2/(4\pi)=1.33$ and $F$ a form factor of the monopole type for the off-shell $\sigma$ with $\Lambda_\sigma=1.7$ GeV, $m_\sigma=550$ MeV. 

In the heavy baryon approximation we can put the external nucleons at rest and, thus, obtain for the elastic scattering of a pion of any charge with a nucleon of any charge
\be
(-it)=2\,\lera{-2i\,\frac{m_\pi^2}{f_\pi^2}\,(c_1^*-c_2^*)}\frac{i}{m_N-m_{N^*}}\lera{iF_\sigma(q_\sigma)g_{\sigma NN^*}}\frac{i}{-m_\sigma^2} \lera{iF_\sigma(q_\sigma)g_{\sigma NN}}\lera{\rho_p+\rho_n}
\label{ropercon}
\ee
where the $\sigma NN$ coupling is  the same as in the Bonn model \cite{Machleidt:1987hj} with $g_{\sigma NN}^2/(4\pi)=5.69$ and $g_{\sigma NN}$ and $g_{\sigma NN^*}$ have the same sign \cite{Hirenzaki:1995js}. 
The contribution in \eqq{ropercon} already contains the sum of the two diagrams on the left hand side of Fig. \ref{fig:roper}.
The isoscalar modification is, thus,
\be
\delta b_0=2\,\frac{(c_1^*-c_2^*) m_\pi^2g_{\sigma NN^*}g_{\sigma NN}m_N}{2\pi f_\pi^2m_\sigma^2(m_N-m_{N^*})(m_\pi+m_N)}\,\rho\simeq 0.184\,m_\pi^{-1}\,\frac{\rho}{\rho_0}.
\label{boroper}
\ee
At normal nuclear matter density $\rho=\rho_0$ this leads to an isoscalar of $\delta b_0=0.188\,m_\pi^{-1}$ which implies attraction. The result from Eq. (\ref{boroper}) is huge compared to the isoscalar from the model of $\pi N$ interaction from Ref. \cite{Doring:2004kt} of $b_c=-0.0336m_\pi^{-1}$, see Tab. \ref{tab:iteration}. 
We can use the $\delta b_0$ from Eq. (\ref{boroper}) and calculate $\Pi_S$ from Eq. (\ref{param_opt}). Then, already at tree level, one obtains the unrealistically large attraction of ${\rm Re}\,\Pi_S=-2.5\cdot 10^{4}\,{\rm MeV}^2$. 

However, by turning the pion line around in the diagram of Fig. \ref{fig:roper}, we have implicitly changed the kinematics at which the above couplings, such as $g_{\sigma NN^*}$, have been determined. The $N^*(1440)$ is now off-shell by around 500 MeV ($E=m_N$) which induces unknown theoretical errors in the calculation. 

Instead of the set $(c_1^*=-7.27\,{\rm GeV}^{-1},\,c_2^*=0\,{\rm GeV}^{-1})$ one can consider the results from Ref. \cite{Meissner:2005bz} which use the combination $c_1^*+c_2^*=(-1.56\pm 3.35)\,{\rm GeV}^{-1}$ from Ref. \cite{Bernard:1995gx} and then apply a resonance saturation hypothesis for the $c^*$ to be saturated by scalar meson exchange. Then the combination $c_1^*-c_2^*$ can be disentangled by the relation $c_1^*/c_2^*=4.2/3.2$ and our result with these values would change from ${\rm Re}\,\Pi_S=-2.5\cdot 10^{4}\,{\rm MeV}^2$ to $(-710\pm 1600)\,{\rm MeV}^2$ at $\rho=\rho_0$ (compare to Figs. \ref{fig:finalnum1}, \ref{fig:finalnum2}). 

\begin{figure}
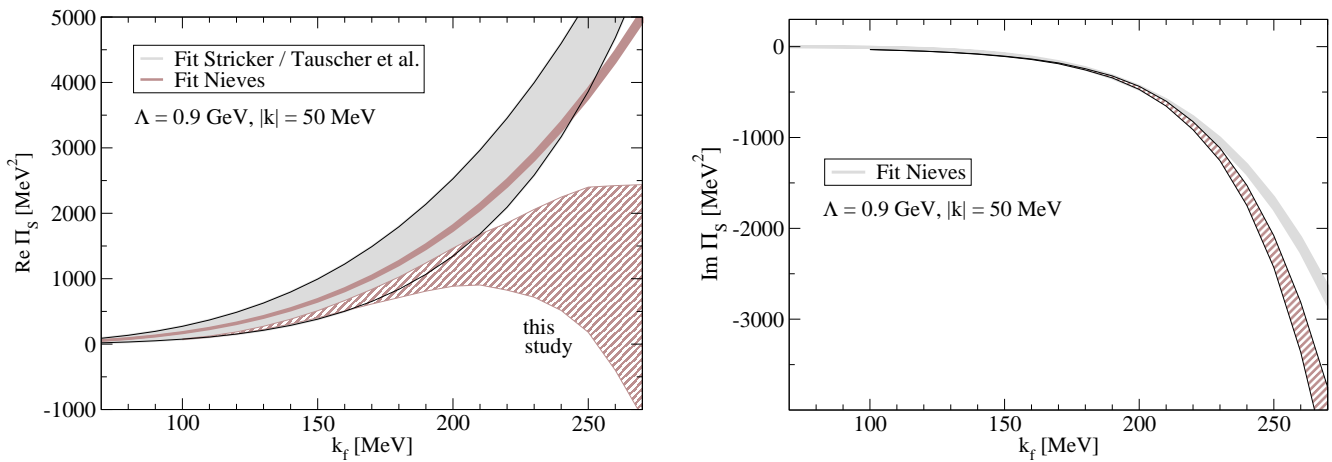

\includegraphics[width=8.5cm]{fig271_op.eps}\hspace*{0.5cm}
\includegraphics[width=8.5cm]{fig272_op.eps}
\caption{Uncertainties of the present study (hatched area). Phenomenological fits to pionic atom data as in Fig. \ref{fig:finalnum1}.}
\label{fig:finalnum3}
\end{figure}

For $\rho=\rho_0/2$, the effective density felt by pionic atoms, the contribution is ${\rm Re}\,\Pi_S=-177\pm 400$ MeV$^2$, which, added to the results calculated before and shown in Fig. \ref{fig:finalnum2}, leads to ${\rm Re}\,\Pi_S=1030\pm 400$ MeV$^2$ for $k^0=m_\pi+10$ MeV (for the Coulomb shift) or ${\rm Re}\,\Pi_S=1430\pm 400$ MeV$^2$ for $k^0=m_\pi+20$ MeV. In both cases this band overlaps with the phenomenological fits, but the amount of theoretical uncertainty is indeed large. In order to account for all the uncertainties, we have taken the results for $k^0=16$ MeV of Coulomb shift, corresponding to an average nucleus with $Z=56$, $A=100$, and have summed in quadrature the uncertainties from the Roper contribution and those of Figs. \ref{fig:c_44_renormalization} and \ref{fig:renoiso}. This leads to the hatched band of theoretical values plotted in Fig. \ref{fig:finalnum3}.
We can see that the band is relatively broad, broader than for ${\rm Im}\,\Pi_S$, since, as we have seen, there are more sources of uncertainty for ${\rm Re}\,\Pi_S$. Altogether, with the realistic theoretical uncertainties accounted for, we find two bands for ${\rm Re}\,\Pi_S$ and ${\rm Im}\,\Pi_S$ which overlap with the empirical values needed to describe pionic atoms, for the effective density $\rho\sim\rho_0/2$ felt by pionic atoms.

\section{Summary and conclusions}
The $s$-wave pion-nucleus optical potential has been calculated in a microscopical many-body approach by looking simultaneously at vertex and selfenergy corrections. We have, thus, taken a chiral unitarized rescattering approach that delivers a good description of vacuum data in the vicinity of the threshold and above. Subsequently, the medium corrections have been added to the vacuum model. Whereas Pauli blocking in rescattering generates repulsion, the pion polarization for intermediate pions, including $ph$, $\Delta h$ and short-range correlations, is responsible for a moderate attraction. The model has been formulated for asymmetric nuclear matter, although vertex renormalizations and other corrections have only been evaluated for symmetric nuclear matter.

For the Weinberg-Tomozawa term and the isoscalar contribution from the NLO chiral Lagrangian, in-medium vertex corrections, some of them novel,  have been included. E.g., the Weinberg-Tomozawa term is increased, in agreement with recent analyses which include also data on deeply bound pionic atoms.

We have also investigated vertex corrections for the NLO isoscalar $\pi N$
amplitude and have found them relevant and of the same relative size as the
renormalization of the isovector vertex. 

When these corrections are taken into account within the multiple scattering
series of the Bethe Salpeter equation, together with selfenergy 
insertions in
the intermediate states and Pauli blocking of the nucleons, we obtain an 
$s$-wave
pion selfenergy, $\Pi_S$, in good agreement with empirical 
determinations for
the the imaginary part and only qualitative for the real part.

 An important ingredient, already exploited in former works, has been 
the effect
 of the Coulomb shift in the argument of $\Pi_S$, which appears in the Klein
 Gordon equation. This effect leads to an increase of the repulsion in
 ${\rm Re} ~\Pi_S$ which brings the pion nucleus optical potential in better 
agreement
 with empirical determinations.
 
  We noticed that an increased $b_1$ in the medium led through 
rescattering to
  an attraction in  ${\rm Re} ~\Pi_S$, in spite of the fact that the Pauli 
correction
  to these terms, given by the Ericson-Ericson formula, produces an 
increased
  repulsion. The rescattering term with the non Pauli blocked part is 
larger than
  with the Pauli blocked one and of opposite sign, thus, altogether,
  rescattering with an increased $b_1$ produces a net attraction. We 
also noted
  that the vacuum model for $\pi N$ has a certain freedom in the choice of
  parameters that induce changes in the real part of the rescattering 
amplitude,
  but  this is compensated by the NLO isoscalar terms in vacuum. We have 
tested
  that, with the simultaneous change in the medium of the isovector 
vertex and
  the NLO isoscalar term, the results for $\Pi_S$ are stable with respect 
to these
  changes of the vacuum model.
 
   Another source of contribution to ${\rm Re} ~\Pi_S$ was a genuine $\rho ^2$ 
term
   related to the Roper coupling to two isoscalar pions. This two 
nucleon term
   was determined, within uncertainties, in former studies of the $\pi d$
   interaction and we have used this information to evaluate the 
contribution
   to ${\rm Re} ~\Pi_S$, resulting in a band of values with relatively large
   uncertainties that one must accept.
  
    Altogether, we determine a band of results for  ${\rm Re} ~\Pi_S$ which 
overlaps
    with empirical determinations in the region of interest of pionic atoms,
    $\rho \sim \rho _0 /2$, and a much narrower band for ${\rm Im} ~\Pi_S$ 
which agrees
    with the also narrower band of empirical analyses in the same region.
   
      Along the work we have pointed out sources of uncertainties which 
we have
   quantified and summed in quadrature at the end. This leads to larger
   uncertainties in ${\rm Re} ~\Pi_S$  than were assumed in former studies, 
and that
   one must bear. Yet, within these admitted uncertainties, the results 
obtained
   represent a satisfactory description of the s-wave pion selfenergy 
for pionic
   atoms.

\end{document}